\documentclass[epj,referee]{svjour}
\usepackage{graphics}
\usepackage{amsfonts}
\usepackage{subfigure}
\usepackage{pgf,pgfarrows,pgfnodes,pgfautomata,pgfheaps,pgfshade}
\usepackage{pslatex}
\usepackage{ae,aecompl}
\usepackage{epsfig}
\usepackage{graphics}
\usepackage{graphicx}
\usepackage{amsmath}
\usepackage{amssymb}
\usepackage{pifont}
\usepackage{psfrag}
\usepackage{subfigure}
\usepackage{float}
\usepackage{color}
\begin{document}
\title{Studies of the limit order book around large price changes}
\subtitle{}
\author{Bence T\'oth\inst{1,2,3}\thanks{E-mail: bence@santafe.edu} \and J\'anos Kert\'esz\inst{3} \and J. Doyne Farmer\inst{1}
}                     
%
%
\institute{Santa Fe Institute, 1399 Hyde Park Road, Santa Fe, NM 87501, USA \and
ISI Foundation - Viale S. Severo 65, 10133 Torino, Italy \and Institute
of Physics, Budapest University of Technology and Economics - Budafoki \'ut. 8. H-1111
Budapest, Hungary}
\date{Received: date / Revised version: date}
%
\abstract{
We study the dynamics of the limit order book of liquid stocks after
experiencing large intra-day price changes.
In the data we find large variations in several microscopical measures, e.g.,
the volatility the bid-ask spread, the bid-ask imbalance, the number of
queuing limit orders, the activity (number and volume) of limit orders
placed and canceled, etc. The relaxation of the quantities is generally very
slow that can be described by a power law of exponent $\approx0.4$.
We introduce a numerical model in order to understand the empirical
results better. We find that with a zero intelligence deposition model of the
order flow the empirical results can be reproduced qualitatively. This
suggests that the slow relaxations might not be results of agents'
strategic behaviour. Studying the difference between the exponents
found empirically and numerically helps us to better identify the role
of strategic behaviour in the phenomena.
\PACS{
      {PACS-key}{discribing text of that key}   \and
      {PACS-key}{discribing text of that key}
     } 
} 
\maketitle
\section{Introduction}\label{intro}

Understanding the relaxation of a system to its typical state after
an extreme event may give a lot of information on the dynamics of
the system. Perhaps the first study of power law relaxations after
extreme events was by Omori who was studying earthquake dynamics \cite{omori1894}.
The Omori law describes the temporal decay of aftershock rates after
a large earthquake: The number of aftershocks is described by a power
law and no typical time scale of the relaxation can be found.
Several further examples can be found for non exponential relaxations
in complex systems: condensed state systems \cite{chamberlin1996}, spin glasses
\cite{bouchaud1992}, microfracturing phenomena \cite{zapperi1997},
internet traffic \cite{johansen2000,abe2003}, etc.
In case of financial markets Ref. \cite{lillo2003} showed that volatility after market
crashes can be also characterised by a power law decay, which is often
referred to as the financial Omori law.

Our aim is to analyse large price changes that can happen
often, maybe every month in case of liquid stocks.
It is important to stress that we are not interested in market crashes or bubbles
(when such changes happen throughout the market) but in large intra-day
price changes specific to a particular stock. Understanding how the market
relaxes after extreme events may be very important for volatility
modeling and forecasting.

Refs. \cite{zawadowski2004,zawadowski2006} analysed the post-event dynamics
of large price changes appearing on short time scales for the New York
Stock Exchange and Nasdaq. They found sharp peaks in the bid-ask
spread, volatility and traded volume at the moment of the events with
slow decay to normal value that in some cases could be characterised
by a power law. Several studies dealt with the analysis of the structure of
the limit order book preceding a large price change \cite{farmer2004,weber2006,ponzi2006,joulin2008}.
Their results show that the volume of market orders play a minor role in
the creation of large price jumps. Instead it is the disappearance of
liquidity in the limit order book that results in extreme price changes.
Ref. \cite{ponzi2006} also studied the relaxation of the bid-ask spread
after large bid-ask spread variations. They found a slow relaxation to normal values,
characterised by a power law with exponent around $0.4$--$0.5$.
Ref. \cite{joulin2008} cross-correlate high-frequency time series of stock
returns with different news feeds, showing evidence that news cannot explain the
price jumps: In general jumps are followed by increased volatility, while
news are followed by lower volatility levels.

In this paper we study the dynamics of the limit order book of the
London Stock Exchange (LSE) around large price changes. Analysing the limit
order book allows us to look at the market at the level of single
orders, and this way connect the microscopic dynamics to macroscopic
measurables and possibly human behaviour. We focus on the post-event
dynamics and the relaxation of the different measures.

We study the dynamics of the volatility and the bid-ask spread near
large events. We find that both have peaks at the moment of the price change.
Their relaxation is slow and can be characterized by a power law.
Analysing the behaviour of market participants, we show results on the
bid-ask imbalance, the number of queuing limit orders in the book,
the aggregated number and aggregated volume of limit orders arriving,
the aggregated number of cancelations and the relative rate of
different order types.  We find that the shape of the
book and the relative imbalance changes very strongly, with a peak at
the event and slow decay afterwards. The activity of both limit orders arriving
and being canceled increases and, after a peak at the event, decays
according to a power law. Surprisingly we find that the relative rates
of limit orders, market orders and cancelations do not vary strongly
in the vicinity of price jumps.

For the relaxation of most of the above measures we find power laws
with similar exponents around $0.3-0.4$. The exponents are very similar,
suggesting that there might be a common cause behind the slow relaxations
of the different measures.

To further study the possible reasons for the relaxation of the volatility and
the bid-ask spread, we construct a zero intelligence multi-agent model
mimicking the actual trading mechanism and order flow. When
introducing large price jumps in the model, we find slow
relaxations in both of the values. This suggests that the slow
relaxations can be reproduced without complicated behavioural
assumptions. We show analytic results on relaxation of the bid-ask spread in
the zero intelligence model.

The paper is built up as follows: In Section \ref{lob} we review the
continuous double auction mechanism and some properties of the limit
order book. In Section \ref{data_methodology} we present the data set
and explain the method of determining large events. Section
\ref{empirical} shows our empirical results. In Section \ref{model}
we present our model and review the
numerical and analytical results. We close the paper by summarising
our results and present the conclusions in Section \ref{conclusions}.

\section{The limit order book}\label{lob}
The market we are studying is governed by a continuous
double auction. Continuous, since orders to buy or sell can be
introduced any time to the market and are matched by an electronic
system in the moment when a match becomes possible.  In the market
agents can place several different types of orders. These can be
grouped into two main categories: \textit{limit orders} and
\textit{market orders}.

Patient traders may submit limit orders to buy or sell a certain
amount of shares of a given stock at a price not worse than a given
limit price. Limit orders are not necessarily executed at the moment
they are submitted. In this case they are stored in the queue of
orders, the limit order book.

On the other hand impatient traders may put market orders, orders to buy
or sell a certain amount of shares of a given stock at the best price
available. Market orders usually are followed by an immediate
transaction, matched to existing limit orders on the opposite side of
the book according to the price and the arrival time.

The third important constituent of market dynamics are cancelations
\cite{eisler2007}:
that is removing an existing limit order from the book. In general,
limit orders increase liquidity, while market orders and cancelations
decrease liquidity.

It is common to analyse \textit{effective} limit and market orders,
i.e. regard all orders that result in an immediate execution as market
orders. This is right from the point of view of the effect of these
orders. However, since we are interested in the behaviour of traders
and in the way their order putting strategies may change, we do not
use this notation, and define limit and market orders by their code in
the limit order book, thus by the intention of the traders.

\subsection{Notations}\label{notations}
Buy limit orders are generally called \textit{bids} and sell limit
orders are generally called \textit{asks}. At any time
instant there exists a buy order with highest price (highest bid), $b_t$,
and a sell order with lowest price (lowest ask), $a_t$, in the
limit order book.
The mean of the best bid and ask prices is the \textit{mid-price}:

\begin{equation}\label{eq:midprice}
m_t=\frac{a_t+b_t}{2}.
\end{equation}

We define volatility as the absolute change of the logarithm of
the mid-price:

\begin{equation}
X_{t}=|\log m_{t}-\log m_{t-1}|,
\end{equation}

\noindent where time is measured in minutes. We apply the absolute 
value of the return for the non-averaged volatility in contrast to its square 
(which leads after averaging to the square of the standard deviation).
The difference between the logarithms of the lowest ask and the
highest bid is called the \textit{bid--ask spread}:

\begin{equation}\label{eq:spread}
S_t=\log a_t- \log b_t.
\end{equation}

\noindent This gives a measure of transaction costs in the market (and on the
other hand the profit of market making strategies).

Further important factors in the limit order book are the \textit{gaps},
i.e. the number of adjacent unoccupied price levels between existing
limit orders. Most often one talks about the first gap, defined as the
difference between the best log-price and the next best log-price in either
side of the book:

\begin{equation}\label{eq:firstgap}
g^{(1)}_{t}=|\log p^{best}_{t} - \log p^{next}_{t}|.
\end{equation}

\noindent The gap is defined in the same way for both sides of the
book and their statistics on the two sides are very similar.
The gap is one tick if adjacent price levels are filled.
The second ($g^{(2)}$), third ($g^{(3)}$), etc. gaps can
be defined in a straightforward way.

\section{Data and methodology}\label{data_methodology}

\subsection{The data set}\label{dataset}
We studied the data of 12 liquid stocks of the LSE
for the period 05.2000 to 12.2002. The stocks studied were:
Astrazeneca (AZN), Baa (BAA), Boots Group (BOOT), British Sky
Broadcasting Group (BSY), Hilton Group (HG.), Kelly Group Ltd. (KEL),
Lloyds Tsb Group (LLOY), Prudential (PRU), Pearson (PSON), Rio Tinto
(RIO), Shell Transport \& Trading Co. (SHEL),
Vodafone Group (VOD).

The LSE consists of two parts: the downstairs market
(SETS) and the upstairs market (SEAQ). The downstairs market is
governed by a fully automatic order matching system, while the
upstairs market is arranged informally through direct connections
between agents. We confine our study to the electronic downstairs market. On the
LSE there are no official market makers, instead every trader can act
as a market maker anytime by posting bid and ask orders
simultaneously. For more details on the rules of the LSE see
\cite{lse}.

\subsection{What are large events?}\label{events}
To study the limit order book dynamics around large events, we first
have to define a filter that determines not only large price
changes but also the moment of the event in a consistent manner. When
determining the events, one has to face the following problem: there
are volatile stocks for which even a price change of 3-4 \% can be an
everyday event, while for some less volatile stocks a much smaller
price change can be the sign of a major event. In filtering for large
events, we are going to follow the method proposed by
\cite{zawadowski2006}.

In order to determine the events, we use two filters on our data.
\begin{enumerate}

\item \textit{Absolute filter} The first filter searches
for intra-day price changes larger than 2 \% of the current price in
time windows not longer than 120 minutes.

\item \textit{Relative filter} We measure the average
intra-day volatility pattern for the stock in the period prior to the
event. The filter searches for intra-day price changes in time windows
not longer than 120 minutes, exceeding 6 times the normal volatility
during that period of the day.

\end{enumerate}

\noindent Normal volatility is defined as the average volatility for
the same period of the day computed over the 60 preceding days.
A price change is considered an event if it passes both of the above
filters. When looking for large events we use transaction prices
(both for returns and volatility) and
the price change is understood to be change in the log-price.
Furthermore we omit the first 5 minutes of the trading day to
exclude opening effects from our measurements. We also omit the last
60 minutes of the trading day in order to be able to study the
intra-day after event dynamics. The method is quite robust when
altering the threshold values.

To be able to localize the exact moment of the events we look at the
earliest and shortest of the time windows in which both of the
thresholds have been exceeded. This means that, when looking for
120 minute events, if the price change already passes the filters in, say, 42
minutes, then we assume the price change has taken place in 42
minutes and the end of the time window is set to the earliest moment
by which the thresholds have been exceeded.
The end of the time window in which the event took place is
considered the end of the price change and from this point we start
studying the post event dynamics. This way the minute 0 of the event
is exactly the end of the time window.

With the filters defined above we were able to determine 289 events
for the 12 stocks in the roughly 2.5-year period. We found a total of
169 downward events and 120 upward events.

\section{Empirical results}\label{empirical}

\begin{figure*}
\centering
\subfigure[][]{%
\label{fig:dyn-vol}%
\includegraphics[width=0.35\textwidth,angle=-90]{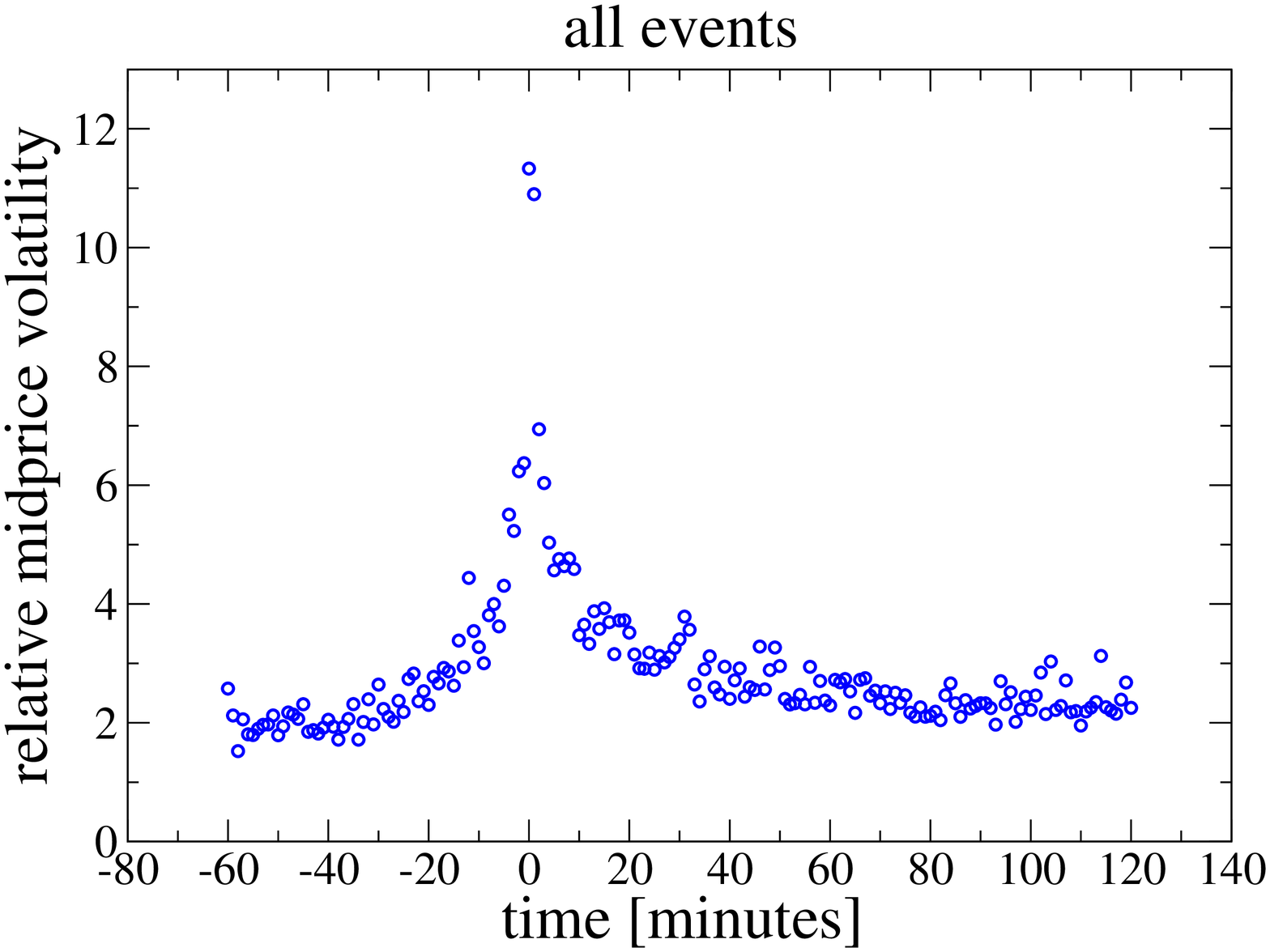}}%
\hspace{8pt}%
\subfigure[][]{%
\label{fig:dyn-spr}%
\includegraphics[width=0.35\textwidth,angle=-90]{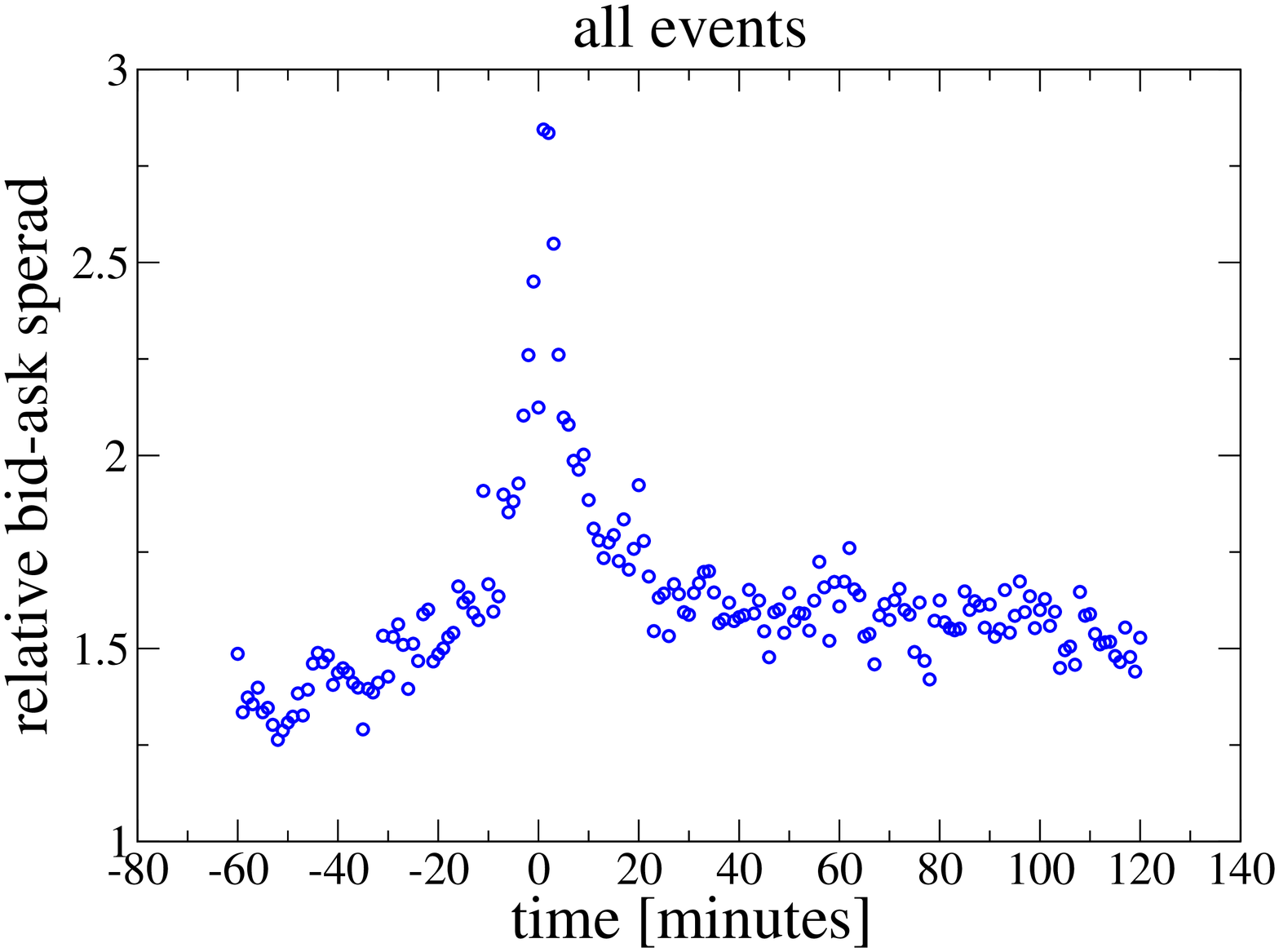}}\\
\subfigure[][]{%
\label{fig:dyn-lo_bid}%
\includegraphics[width=0.35\textwidth,angle=-90]{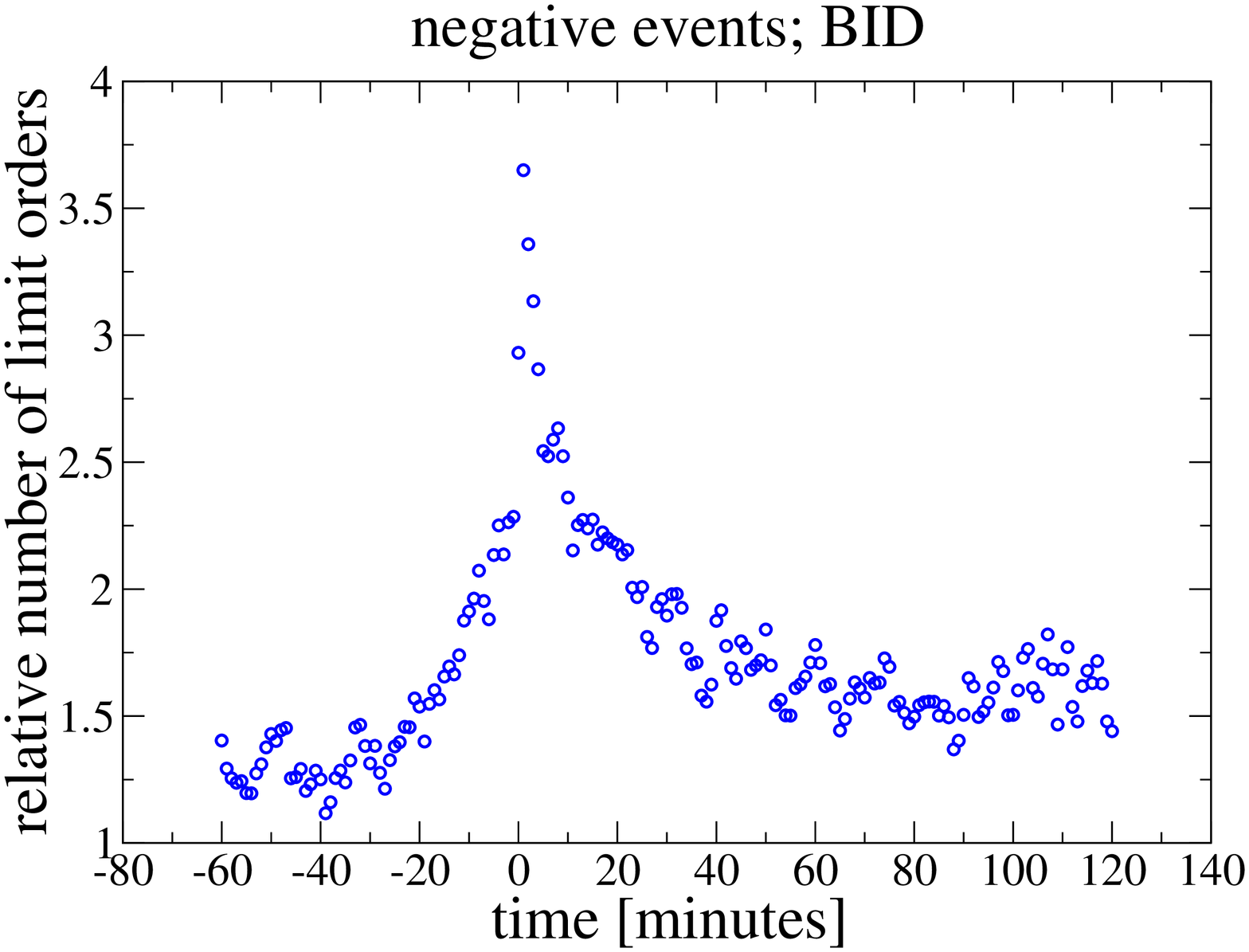}}%
\hspace{8pt}%
\subfigure[][]{%
\label{fig:dyn-lo_ask}%
\includegraphics[width=0.35\textwidth,angle=-90]{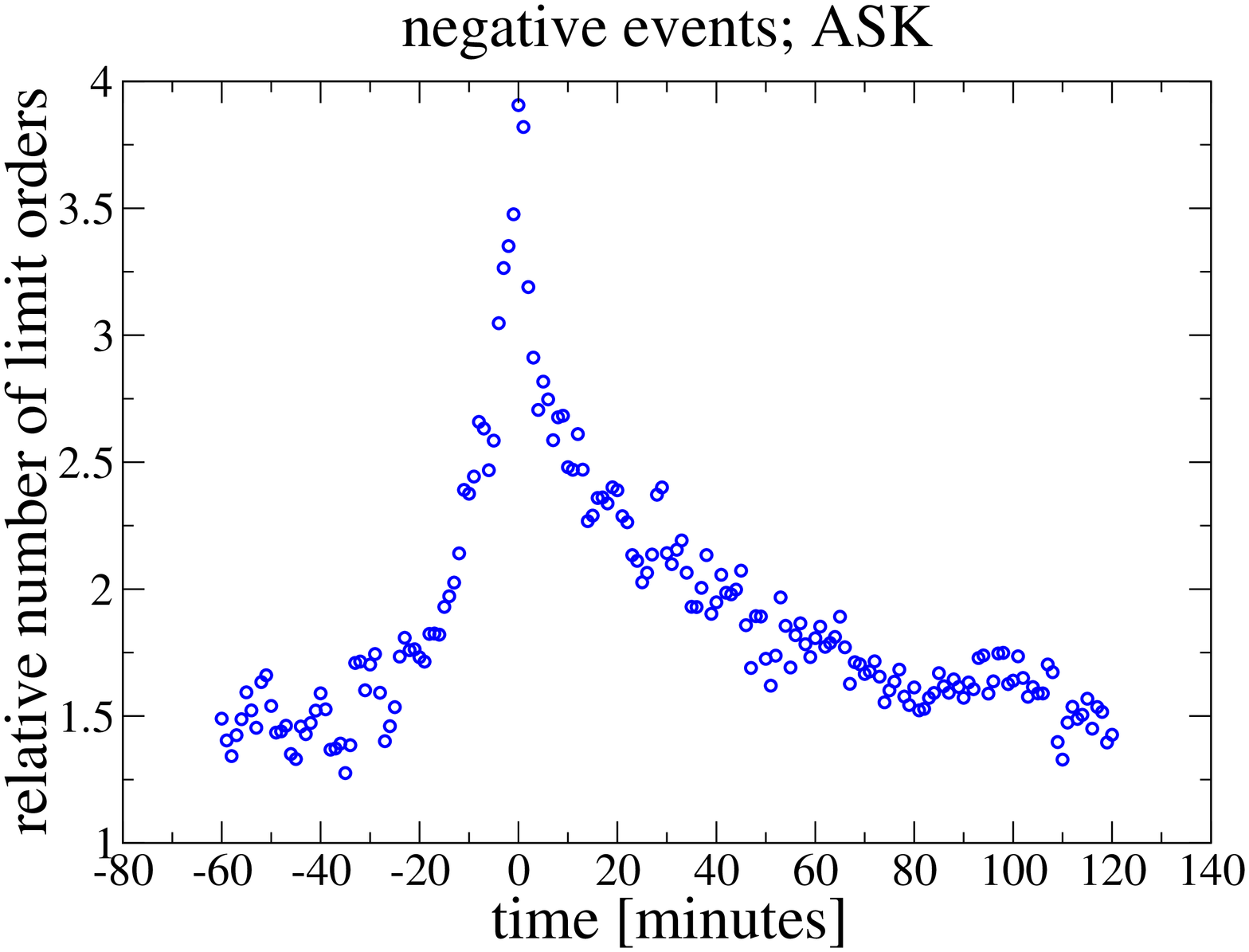}}\\
\subfigure[][]{%
\label{fig:dyn-c_bid}%
\includegraphics[width=0.35\textwidth,angle=-90]{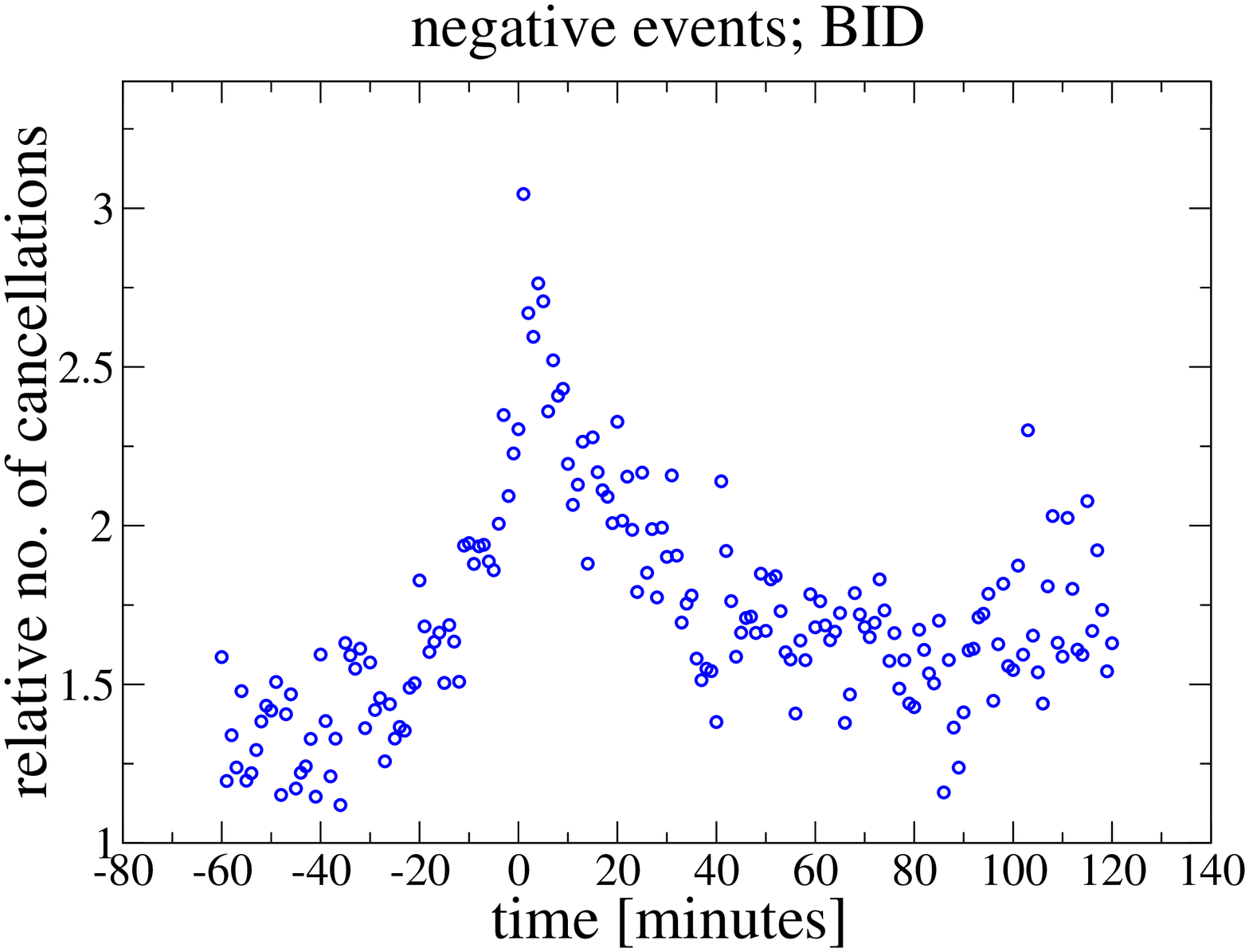}}%
\hspace{8pt}%
\subfigure[][]{%
\label{fig:dyn-c_ask}%
\includegraphics[width=0.35\textwidth,angle=-90]{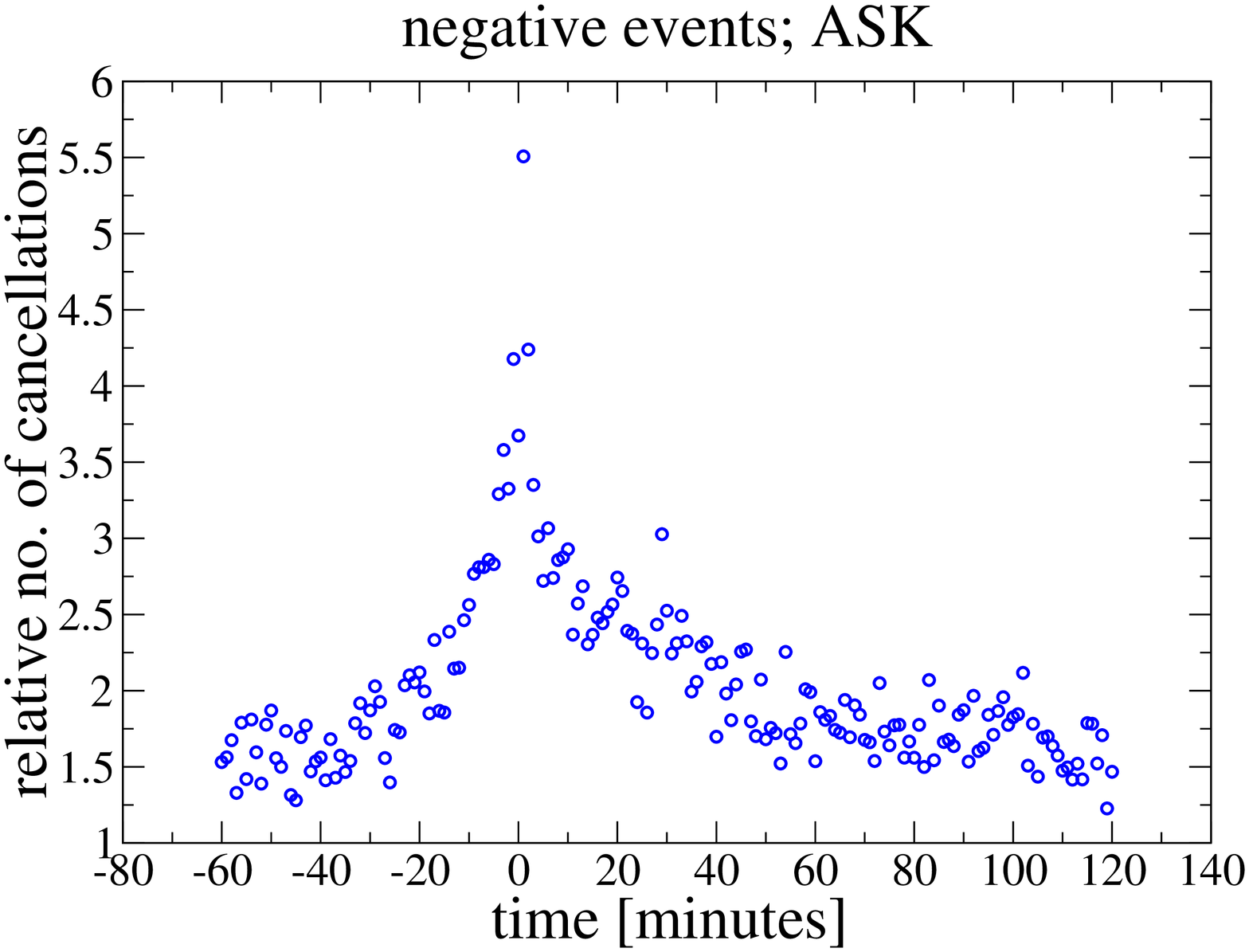}}\\
\caption[A set of four subfigures.]{(Color online) Dynamics around events compared to the usual
dynamics. Usual dynamics are computed from the 60-day pre-event average for the
same period of the day.
\subref{fig:dyn-vol} Mid-price volatility averaged over all events;
\subref{fig:dyn-spr} Bid-ask spread averaged over all events;
\subref{fig:dyn-lo_bid} The number of limit orders in case of negative events on the bid side of the book and
\subref{fig:dyn-lo_ask} on the ask side of the book;
\subref{fig:dyn-c_bid} The number of limit order cancelations in case of negative events on the bid side of the book and
\subref{fig:dyn-c_ask} on the ask side of the book.}%
\label{fig:dyn}%
\end{figure*}

We studied the dynamics of several measures of the order book.
In Figure \ref{fig:dyn} we show the changes in the volatility, the bid-ask
spread and the number of limit orders placed and canceled. All plots show the
dynamics for the 60-minute pre-event and 120-minute post-event periods.
For all events we defined time 0 as the end of the shortest and earliest time windows in
which the filters defined in Section \ref{events} have been passed. Then, in
case of each event we compared the dynamics in the period from -60
minutes to 120 minutes to the average volatility dynamics for the same
period of the day computed over the 60 days preceding the event. In
the next step we aggregated the dynamics over the events. This way
all the plots show dynamics compared to the average dynamics, daily periodicities
excluded.

Figure \ref{fig:dyn-vol} shows the dynamics of the volatility and
Figure \ref{fig:dyn-spr} shows the dynamics of the bid-ask spread. In this
case we included both downward and upward price jumps in our sample.
Figure \ref{fig:dyn-lo_bid} and \ref{fig:dyn-lo_ask} show
how the number of limit orders placed has changed on the bid side and
the ask side of the book in case of downward price jumps.
Figure \ref{fig:dyn-c_bid} and \ref{fig:dyn-c_ask} show
how the number of limit orders canceled has changed on the bid side and
the ask side of the book in case of downward price jumps.
The number of orders placed and canceled may be regarded as measures of the \
activity of market participants.
(In case of the activities, the plots for upward price jumps are
very similar to those with downward jumps, so we only show the negative events.)

\begin{figure*}
\centering
\subfigure[][]{%
\label{fig:relax-vol}%
\includegraphics[width=0.35\textwidth,angle=-90]{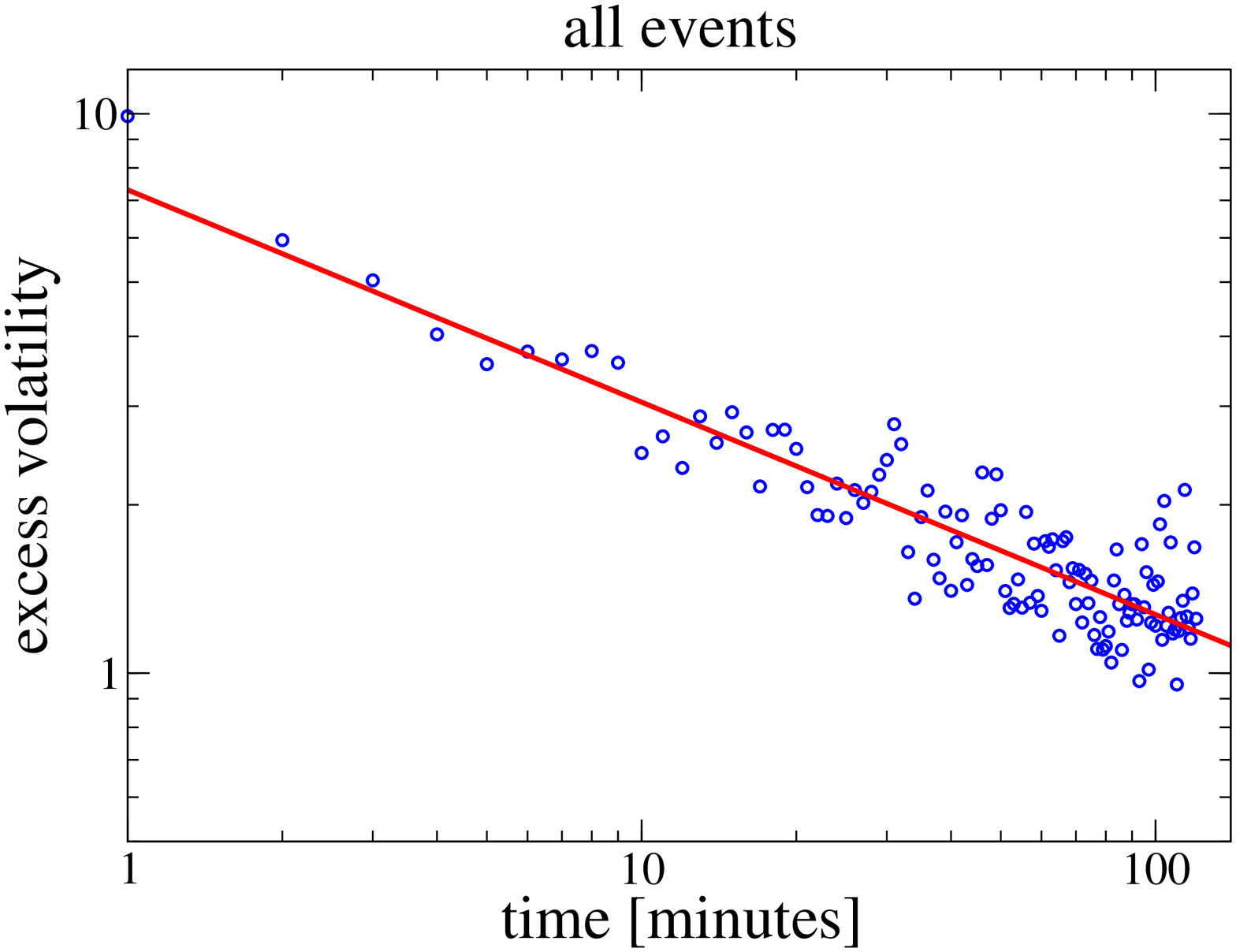}}%
\hspace{8pt}%
\subfigure[][]{%
\label{fig:relax-spr}%
\includegraphics[width=0.35\textwidth,angle=-90]{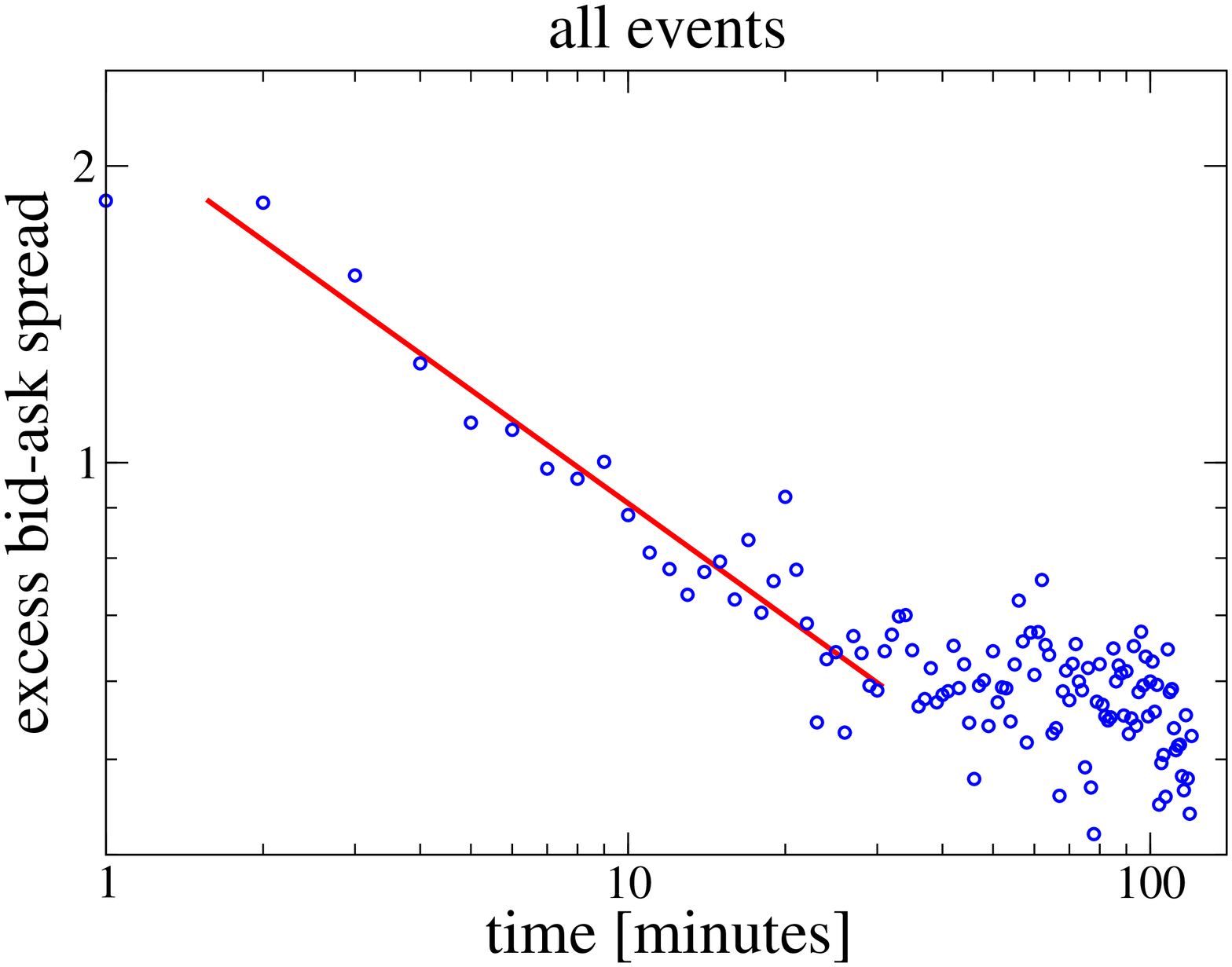}}\\
\subfigure[][]{%
\label{fig:relax-lo_bid}%
\includegraphics[width=0.35\textwidth,angle=-90]{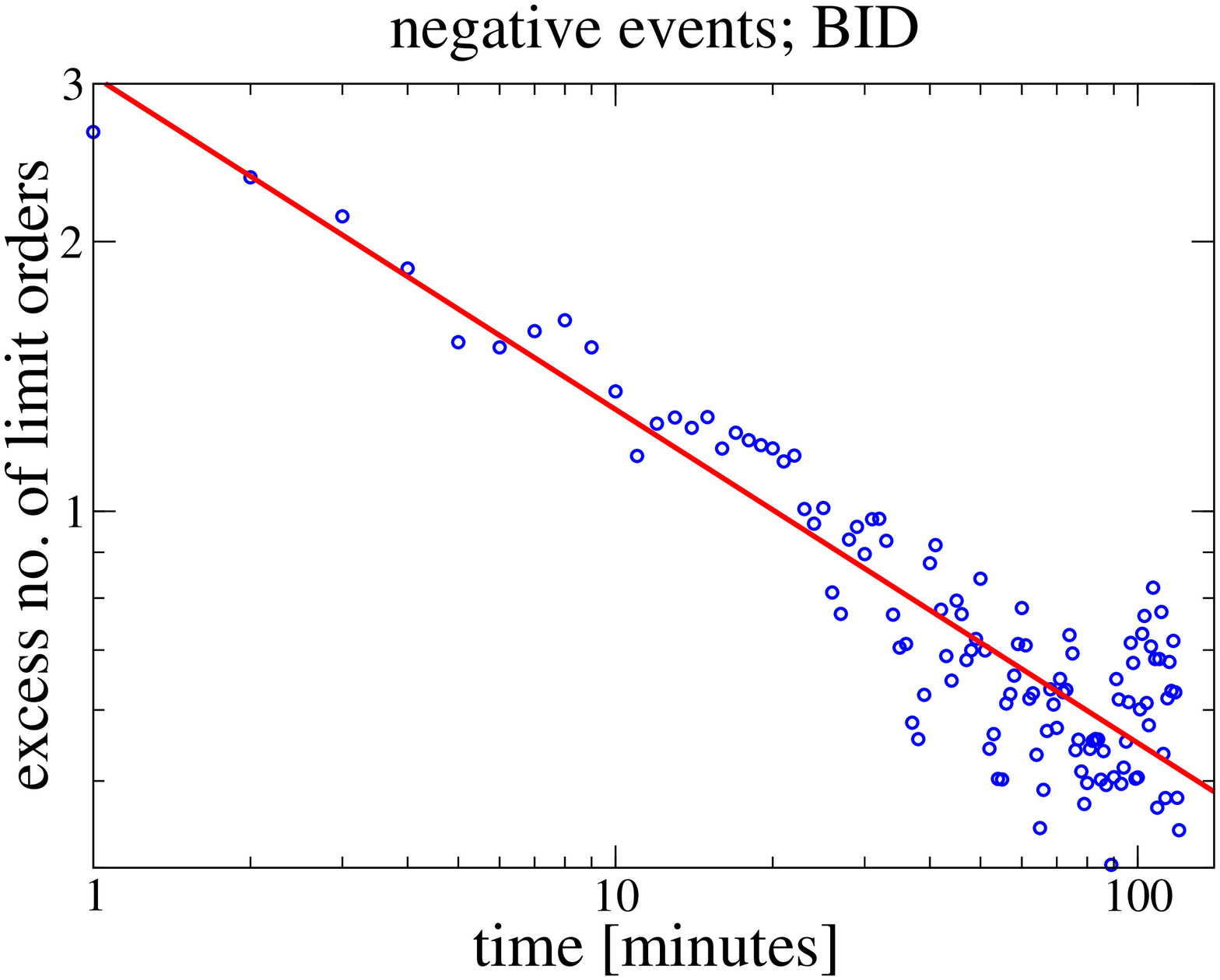}}%
\hspace{8pt}%
\subfigure[][]{%
\label{fig:relax-lo_ask}%
\includegraphics[width=0.35\textwidth,angle=-90]{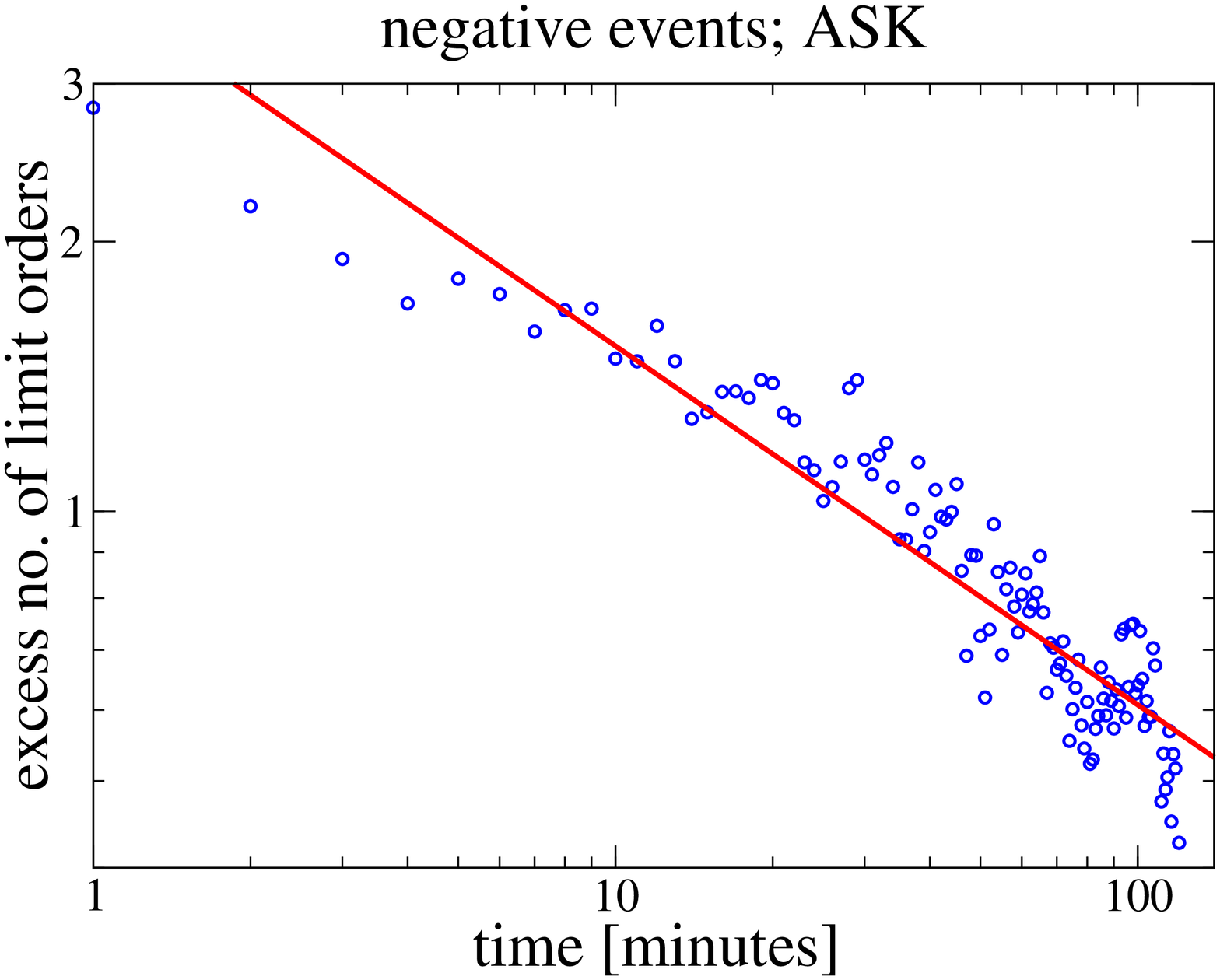}}\\
\subfigure[][]{%
\label{fig:relax-c_bid}%
\includegraphics[width=0.35\textwidth,angle=-90]{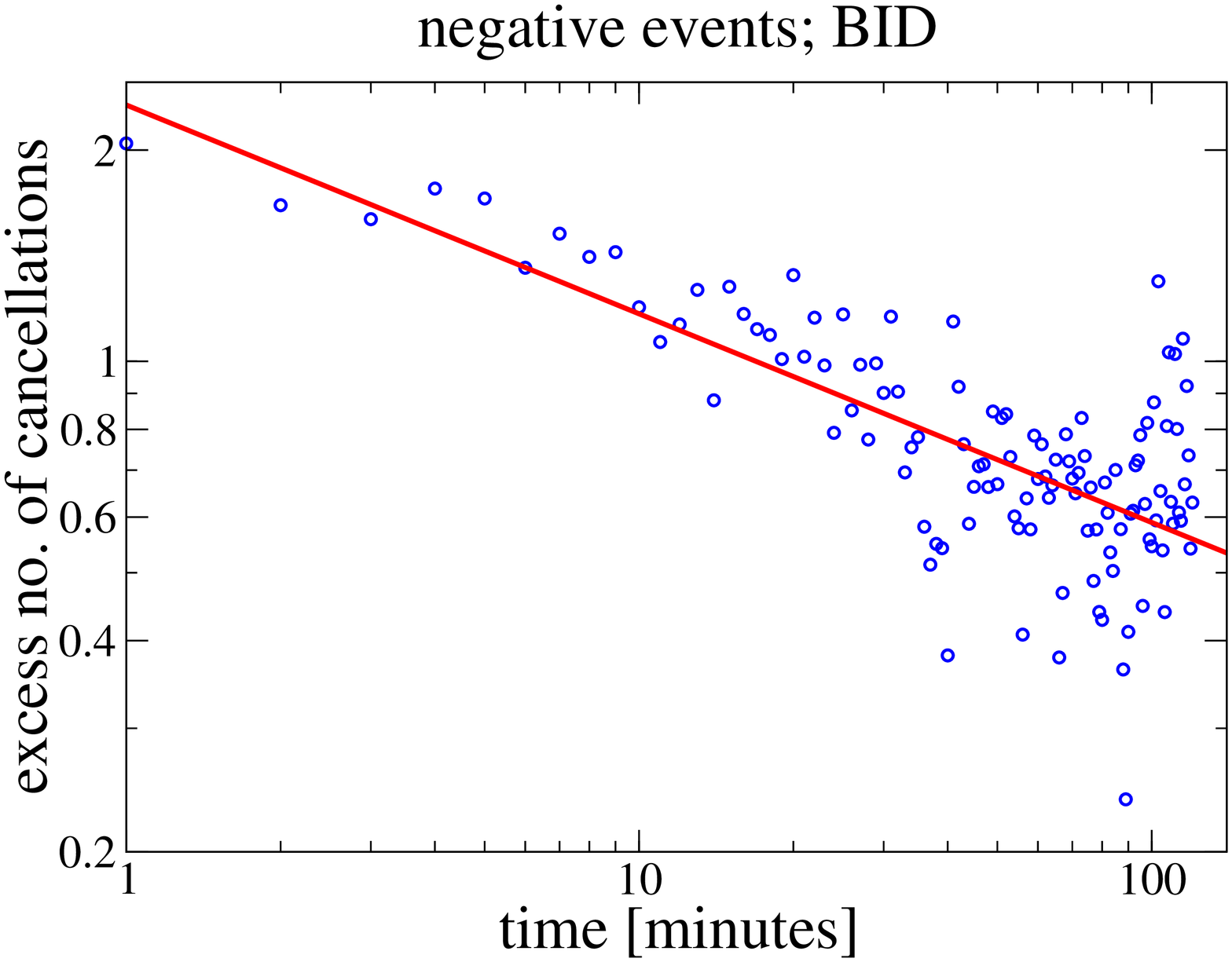}}%
\hspace{8pt}%
\subfigure[][]{%
\label{fig:relax-c_ask}%
\includegraphics[width=0.35\textwidth,angle=-90]{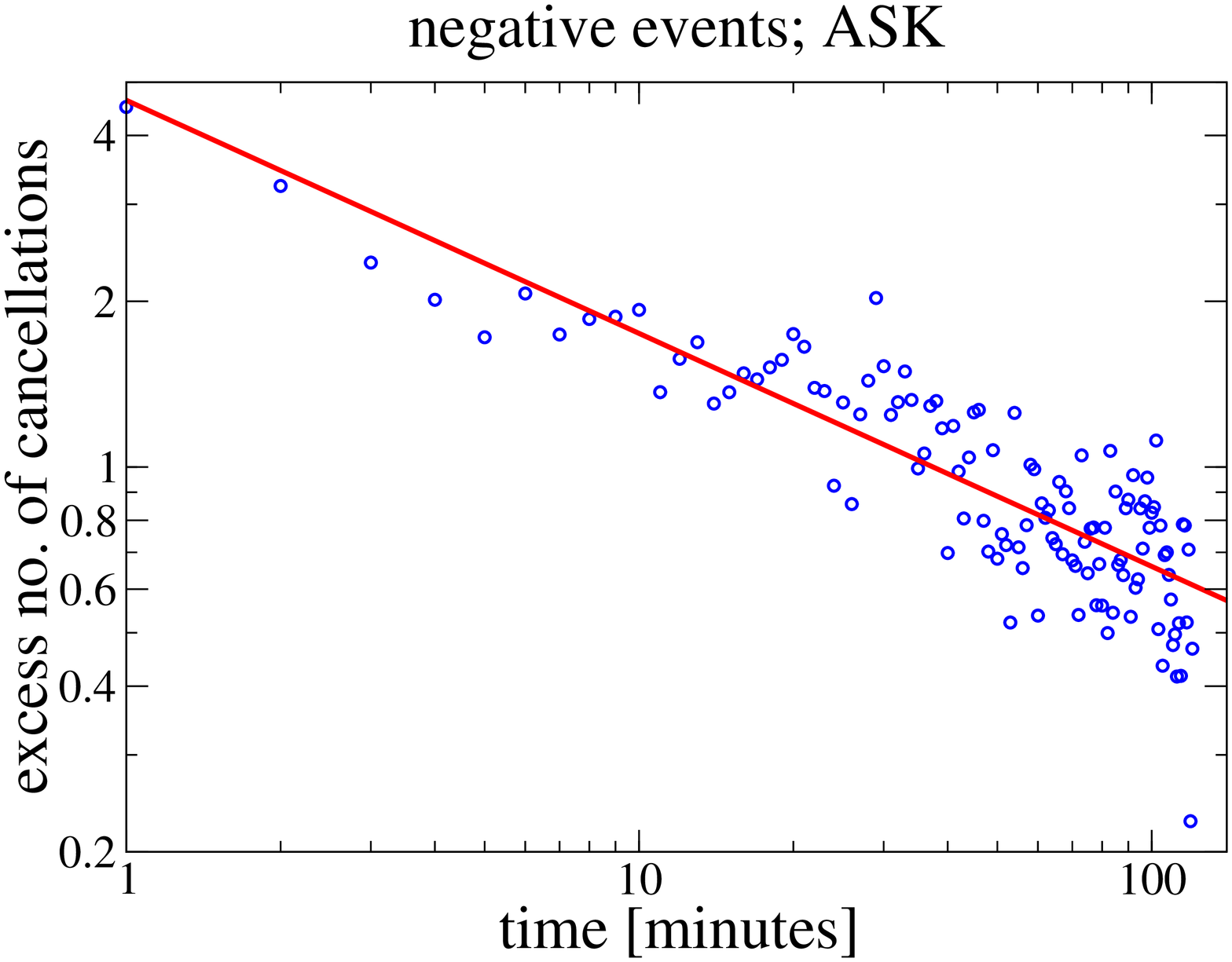}}\\
\caption[A set of four subfigures.]{(Color online) Relaxation of the excess variables on log-log scale with
power law fits:
\subref{fig:relax-vol} Mid-price volatility averaged over all events, exponent: $0.38\pm 0.01$;
\subref{fig:relax-spr} Bid-ask spread averaged over all events. The first part of the relaxation curve seems to show scaling behaviour and may be fit by a power law with an exponent: $0.38\pm 0.03$. However the scaling for the bid-ask spread is less consistent than that
of the volatility. Also the long time decay is apparently even slower than
a power law.;
\subref{fig:relax-lo_bid} The number of limit orders in case of negative events on the bid side of the book, exponent: $0.37\pm 0.01$, and
\subref{fig:relax-lo_ask} on the ask side of the book, exponent: $0.40\pm 0.01$;
\subref{fig:relax-c_bid} The number of limit order cancelations in case of negative events on the bid side of the book, exponent: $0.30\pm 0.02$, and
\subref{fig:relax-c_ask} on the ask side of the book, exponent: $0.42\pm 0.02$.}%
\label{fig:relax}%
\end{figure*}

We find that the dynamics are very similar. All measures have a strong peak
at the moment of the event determined by our filter. We are going to focus
on the relaxation of the variables after the peak: This is the part that
is well defined by our filtering method.
Concerning the \emph{rise} prior to the events we have to be
careful, however. Due to our method of determining events (the fact that
we define the end of the time window as zero time of the event) the pre-event
dynamics are conditioned on the event and can not be regarded as independent.
Practically, what we can say is that the variables change near the event
with a peak at the moment of the event, but the actual rise can not be
characterised through these results.
The relaxations after the event seem to be slow in all cases.
In Figure \ref{fig:relax} we show the relaxation of the same measures after
the event on a log-log scale.
In order to quantify the relaxations we apply power
law fits to the curves and compare the power law exponents.
(Note that to study the relaxations to the normal
value, we plot the excess variables, thus the difference between the actual
value and the value in normal periods.)
In Table \ref{table:exponents} we summarise the exponents of the relaxations.

As we can see the \emph{volatility} increases very high, to roughly 12 times its
normal value and its decay can be well fit by a power law with an exponent
of approximately 0.38.
The result on the power law relaxation with exponent 0.38
of the excess volatility after events can be compared to results on the
relaxation after crashes. Ref. \cite{lillo2003} shows that volatility
after stock market crashes decays approximately with an exponent of
0.3, showing that the post-crash dynamics are similar to those of
earthquakes, commonly known as the Omori law. Another study
showed that the relaxation of excess
volatility after fluctuations is characterized by a power law with
exponent between 0.3-0.4, measured to be robust for different time
periods and across markets \cite{zawadowski_private}. This may point us to the fact that
relaxation of the volatility after extreme events and after
fluctuations are similar, also mentioned in \cite{weber2007}.

The \emph{bid-ask spread} increases, but to a lower value than the volatility,
with a peak of roughly 3 times the average. The first part of the relaxation curve
seems to show scaling behaviour and
may be fit by a power law. Interestingly, we find that the exponent of the decay is
around 0.38, very close to the exponent of the volatility decay. The
slow, power law like decay of excess bid-ask spread is in agreement with the
findings of \cite{ponzi2006}.
However the scaling for the bid-ask spread is less consistent than that
of the volatility. Also the long time decay is apparently even slower than
a power law.
Interestingly, \cite{zawadowski2006} found strong variation of the
bid-ask spread after large events at NYSE, but not on
Nasdaq. They argued that this difference is caused
by the diverse trading rules on the two markets, pointing out that the
existence of a single market maker at NYSE leads to the opening of the
bid-ask spread. The trading rules of LSE, being a fully automatic market makes
it similar to Nasdaq, so, according to that argument, one would
expect low variation of the bid-ask spread on the LSE. This is not consistent
with our findings. An explanation of this contradicting behaviour on the
two markets is missing at present.

The \emph{number of limit orders placed} increases both on the bid
and the ask side of the book. We find a peak of roughly four
times the usual value with a slow decay of the activity afterwards.
The relaxation of the excess limit order putting activity after events
can be described by a power law decay with an exponent of roughly 0.37--0.4.
Concerning the dynamics of the aggregated \emph{volume of
limit orders placed} we get very similar results (not shown here). However,
in case of the volumes
we see a slightly stronger increase around the event with roughly
6 times the usual value. This suggests that the average volume of
limit orders increases around large price changes.
The relaxation of the excess limit order volume can be characterised
by a power law of exponent roughly 0.44 in case of negative events and
exponent 0.48 for positive events.

Similarly, the \emph{number of limit orders canceled} increases
strongly around events, with a slow decay after the peak.
It also seems that the increase is
stronger and much more clear on the side of the book which is
opposite to the direction of the price change, i.e. ask side in case
of negative events and bid side in case of positive events.
For the relaxations we find that in general, the decays on the
opposite side of the book compared to the direction of the price
change can be well fit by a power law of exponent roughly 0.42. The
decays on the same side as the price change direction are more noisy
but may be fit by a power law, showing exponents of roughly
0.3--0.35.

\begin{table}
\centering
\caption{The exponents of the relaxation curves for the different variables.}
\label{table:exponents}       
\begin{tabular}{ll}
\hline\noalign{\smallskip}
variable & exponent\\
\noalign{\smallskip}\hline\noalign{\smallskip}
volatility & $0.38\pm 0.01$\\
bid-ask spread & $0.38\pm 0.03$\\
limit orders placed - bid & $0.37\pm 0.01$\\
limit orders placed - ask & $0.40\pm 0.01$\\
cancelations - bid & $0.30\pm 0.02$\\
cancelations - ask & $0.42\pm 0.02$\\

\noalign{\smallskip}\hline
\end{tabular}
\vspace*{5cm}  
\end{table}

It is generally understood that the relative amount of supply and demand
govern the movement of prices. To quantify the pressure of
orders from either side of the book, we study the \emph{bid-ask imbalance} on
the market. We denote the total volume to buy on the market by
$V_{t}^{buy}$ and the total volume to sell by $V_{t}^{sell}$. Then we
define the buy imbalance and the sell imbalance as:

\begin{equation}
I_{t}^{buy}=\frac{V_{t}^{buy}}{V_{t}^{buy}+V_{t}^{sell}}
\end{equation}

\noindent and

\begin{equation}
I_{t}^{sell}=\frac{V_{t}^{sell}}{V_{t}^{buy}+V_{t}^{sell}}.
\end{equation}

\noindent Trivially, $I_{t}^{buy}+I_{t}^{sell}\equiv 1$.

\begin{figure*}[t!]%
\centering
\subfigure[][]{%
\label{fig:imb-up}%
\includegraphics[width=0.35\textwidth,angle=-90]{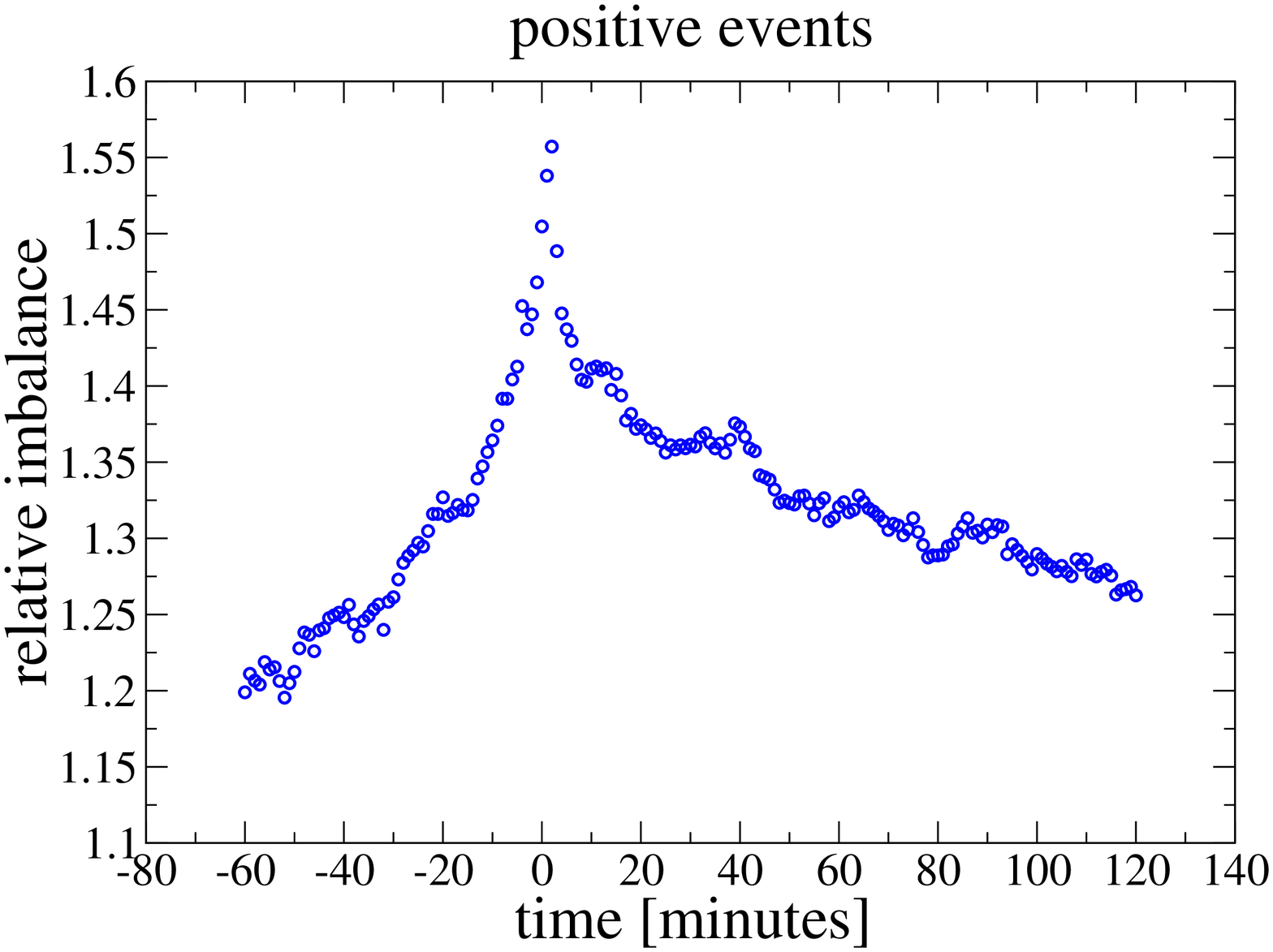}}%
\hspace{8pt}%
\subfigure[][]{%
\label{fig:imb-down}%
\includegraphics[width=0.35\textwidth,angle=-90]{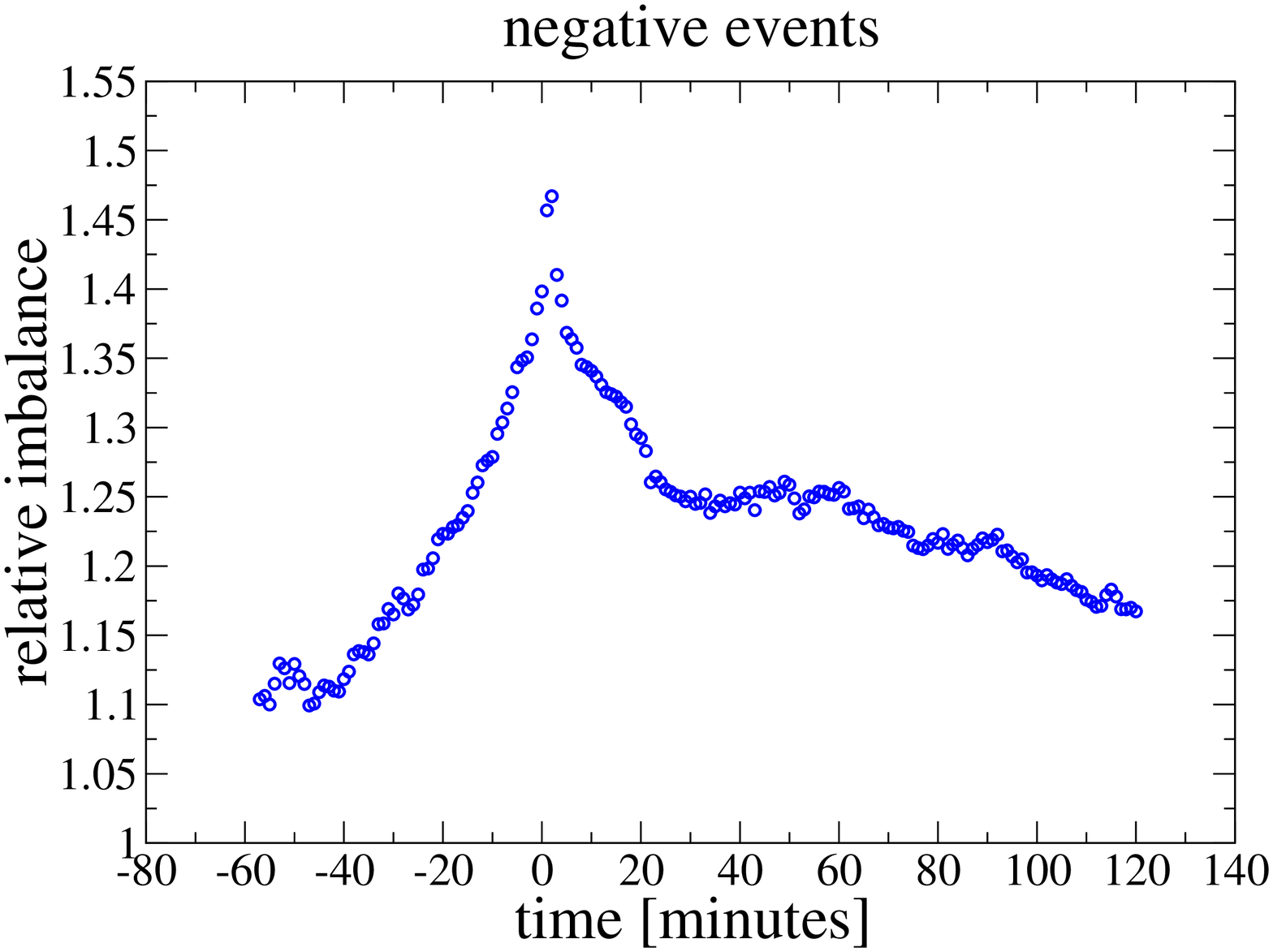}}\\
\caption[A set of four subfigures.]{(Color online) The dynamics of the
imbalance in the volume of supply and demand:
\subref{fig:imb-up} shows the buy imbalance ($I^{buy}$) in case of upward price jumps;
\subref{fig:imb-down} shows the sell imbalance ($I^{sell}$) in case of downward price jumps.}
\label{fig:imb}%
\end{figure*}

In Figure \ref{fig:imb} we show the dynamics of the imbalances
for upward and downward price changes,
compared to the 60 day pre-event interval for the
same period of the day, as before. The two curves are very similar
showing a vanishing amount of orders to sell in case of upward price moves
and vanishing amount of orders to buy in case of downward price moves.

To understand the absolute values on the y-axis
is not straightforward, to make it clear we show the following: If
we assume that in regular market periods, on average half of the
volume of orders is to buy and consequently half is to sell, then the
values on the y-axis show that in case of the moment of a
positive event, only $\approx 100\%-50\%*1.57=21.5\%$ of the total
volume of orders appears on the sell side of the book and
in case of the moment of a negative
event, only $\approx 100\%-50\%*1.48=26\%$ of the total volume of orders
appears on the buy side of the book. These numbers
show that there is a huge imbalance of volume around events. The
relaxation of the imbalance after the events is very slow.

\begin{figure*}[t!]%
\centering
\subfigure[][]{%
\label{fig:queue_down}%
\includegraphics[width=0.35\textwidth,angle=-90]{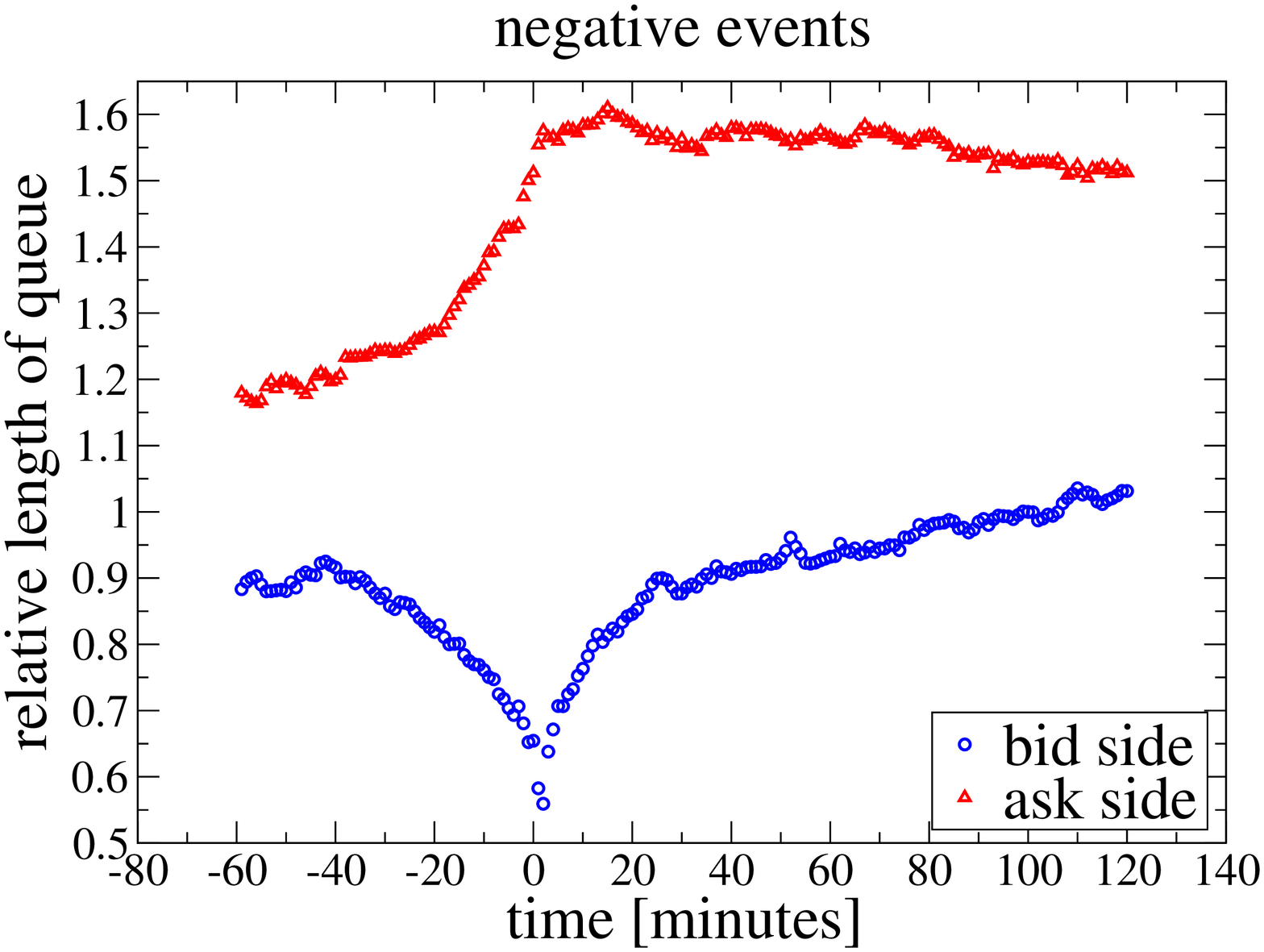}}%
\hspace{8pt}%
\subfigure[][]{%
\label{fig:rates_down}%
\includegraphics[width=0.35\textwidth,angle=-90]{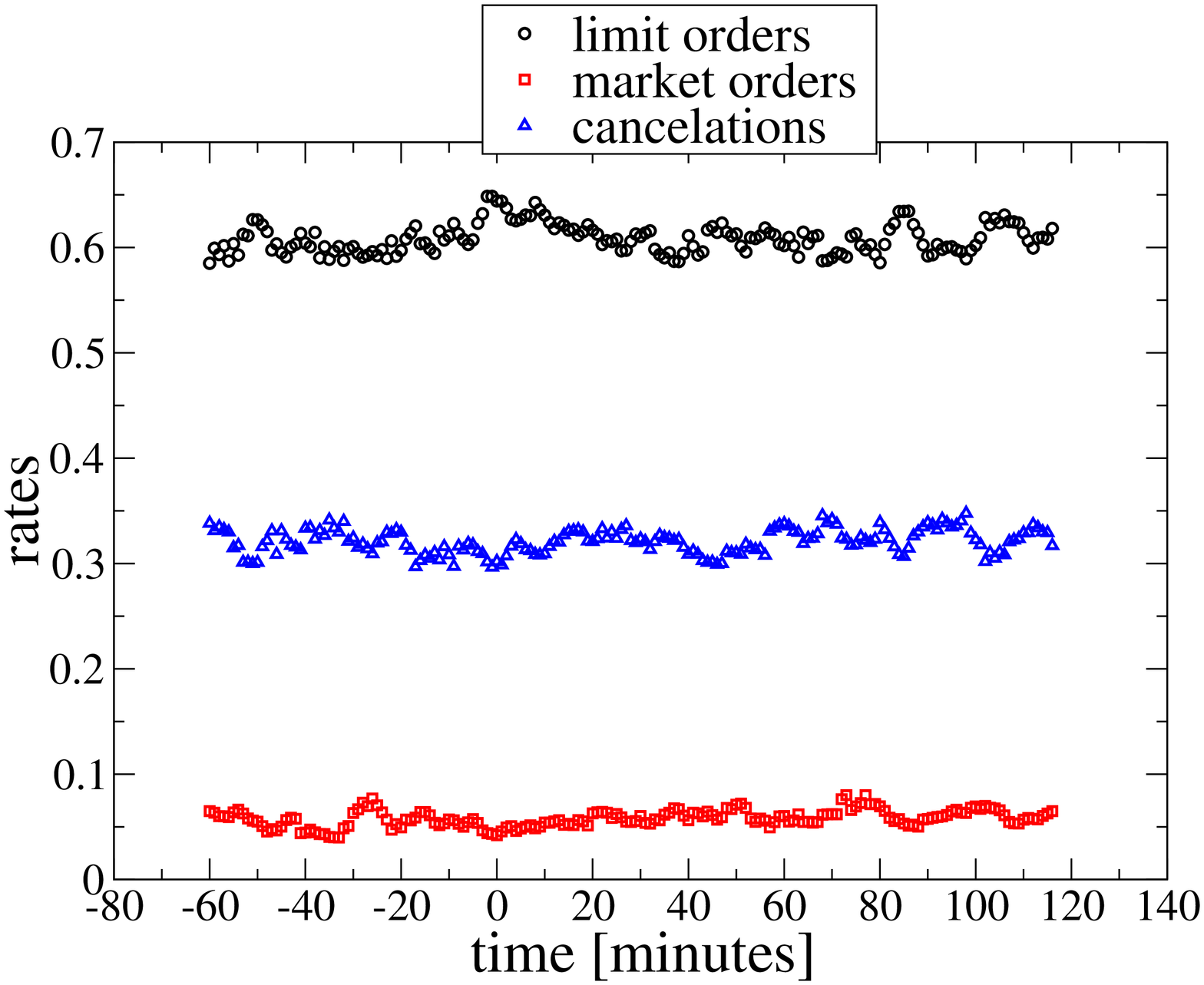}}\\
\caption[A set of four subfigures.]{(Color online)
\subref{fig:queue_down} Dynamics of the number of queuing orders
in the book in case of negative events. The number of queuing limit
orders on the bid side decreases to about half of the usual value and
only relaxes back slowly. The number of queuing limit orders on the
ask side increases to roughly 1.6 times the usual value and even after
the event stays very high for a long time.;
\subref{fig:rates_down} The dynamics of the rates of limit orders
(black circles), market orders (red squares) and cancelations (blue
triangles) around negative events for the buy side of the book.
No very strong variations can be
seen in the different rates around the large price changes.}
\label{fig:mixed}%
\end{figure*}

A measure very similar to the bid-ask imbalance, is the \emph{number
of queuing orders} on either side of the limit order book.
Figure \ref{fig:queue_down} shows the dynamics of the number of
queuing orders in the limit order book compared to the 60 day
pre-event interval for the same period of the day, for negative
events. The plot shows results for both the bid and the ask sides of
the book. The results for positive events are very similar (of course
symmetrically to the case of negative events), so we only show the
figures for the negative events.

We can see that for negative price jumps the number of queuing limit
orders on the bid side decreases to about half of the usual value and
only relaxes back slowly. At the same time on the ask side the number
of queuing orders increases to roughly 1.6 times the usual value and
even after the event stays very high for a long time. Further studies
are needed to see if this very slow relaxation is a sign of the
limit order book being partly frozen in post-event periods.

It is interesting to study the rates of different market activities,
i.e. the relative number of limit orders, market orders and
cancelations, compared to the total number of orders. Figure \ref{fig:rates_down}
presents the dynamics of the three rates around negative events for
the buy side of the book.
We see that there are no strong changes in the rate
around large price changes. The results for the ask side of the book are the same.
When studying positive events we get very
similar dynamics: No strong variation in the relative rates.
This result means that it is rather the entire market activity
changing (increasing) in the surroundings of large events and not the
strategy of traders how they place orders.\footnote{As we
stated before, we regard orders as limit or market orders by the
intention of the trader, not by their effect (i.e. not effective limit
and market orders). Because of this we get a higher rate of limit
orders and lower rate of market orders, than presented in
\cite{farmer2004}. However the rate of cancelations fits their
results very well, so we believe that after accounting for the
effective orders, there are no contradictions between the two results.}

Summarising the empirical results: We have studied the dynamics of
several measures of
the order book before and after large price changes.
For the change in the volatility, the bid-ask spread, the limit order
placing and cancelation activity,
the bid-ask imbalance, the number of queuing orders in the book
we found strong variation at the moment of the event and slow
relaxation in the post-event
period. Specially, in case of the volatility, the bid-ask spread and the
activities we found a relaxation very similar to a power law, with
exponents close to 0.4 suggesting a possible common cause behind
the slow relaxations of the different measures.

Analysing the rates of limit orders, market orders and cancelations
around large events, we did not find strong variation, showing that
strategy of traders in choosing their type of order does not
vary much.

\section{An agent-based model}\label{model}
As stated above, we found very similar relaxation in different
measures of the limit order book.  To better understand the dynamics
leading to the slow relaxations, in this section we introduce a
multi-agent model of the order placing and removing process. When
constructing a modeling framework, we have to decide which path to
follow:

\begin{enumerate}
\item Building a multi-agent model with complicated strategies,
involving behavioural assumptions.
\item Building a \textit{zero intelligence} multi-agent model,
with some very basic assumptions on the order flow.
\end{enumerate}

\noindent In the literature there are several examples for both types of
models. Models of type 1 permit one to study behavioural results of
the model, but when building the trading strategies we have to be
careful, not to assume unrealistic properties of traders and/or avoid
the common error, to simply find as output exactly the input
assumptions. Models of type 2 are easier to construct, but apart
from the problem of possible over-simplification, they also confine us
to the analysis of non-behavioural measures through the
model.

We chose to follow the path of
\cite{stigler1963,bak1997,maslov2000,challet2001,willmann2002,muchnik2003,daniels2003,smith2003}
to construct a
zero intelligence multi agent model of the continuous double auction
through the limit order book. The model is aimed to be as simple as
possible but capturing the most important properties of the continuous
double auction. If we are able to reproduce some results with a zero intelligence model
it may suggest that the particular phenomenon is not due to
traders' strategic behaviour but rather to the market mechanism or institutions.

\subsection{Details of the model}\label{model_details}

We assume limit order placing and cancelations similarly to a
deposition--evaporation process and furthermore introduce market
orders.
Our model is similar to Maslov's model \cite{maslov2000}.
The main differences are that we allow for cancelation of
existing limit orders (and through this the tuning
of the probabilities of different actions) and that agents put
their limit orders relative to the mid-price, this way we allow
for a non trivial dynamics of the bid-ask spread.
All orders arrive or evaporate with the same unit volume. The
trading mechanism is the following:

\begin{itemize}
\item Limit orders arrive with rate $P_{LO}$ per unit time with equal
probability to buy or sell. Limit orders get deposited in the
interval $[m_{t}-D,m_{t}]$ in case of buy orders and in the interval
$[m_{t},m_{t}+D]$ in case of sell orders with uniform distribution,
where $m_{t}$ is the mid-quote price (see Equation
\ref{eq:midprice}) and $D$ is a parameter of the model.
\item Market orders arrive with rate $P_{MO}$ per unit time with equal
probability to 'buy' or 'sell'. A market order to buy (sell) will get executed
immediately by being matched to the best limit order to sell (buy).
\item Existing limit orders are being canceled with rate $P_{C}$ per
unit time from the 'buy' or 'sell' side with equal probability. In
case of a cancelation on one side of the book, all limit orders on
that side have the same probability: $p=V_{total}^{-1}$ to be
canceled where $V_{total}$ is the total volume of limit orders on
that side of the book. Thus on average one limit order evaporates from
the book in case of cancelations.
\end{itemize}

\begin{figure}[htb!]
\resizebox{0.25\columnwidth}{!}{%
\vbox{
\includegraphics{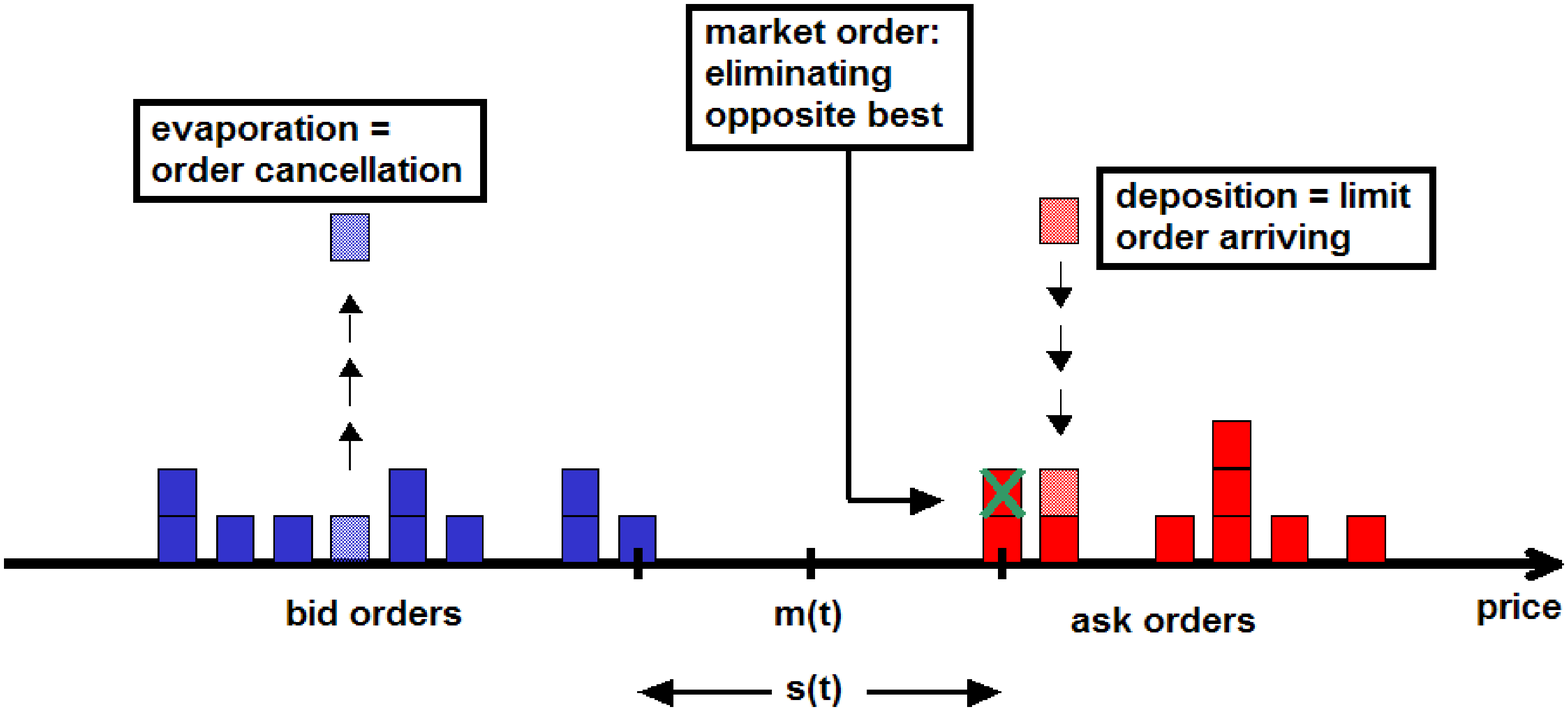}
\includegraphics{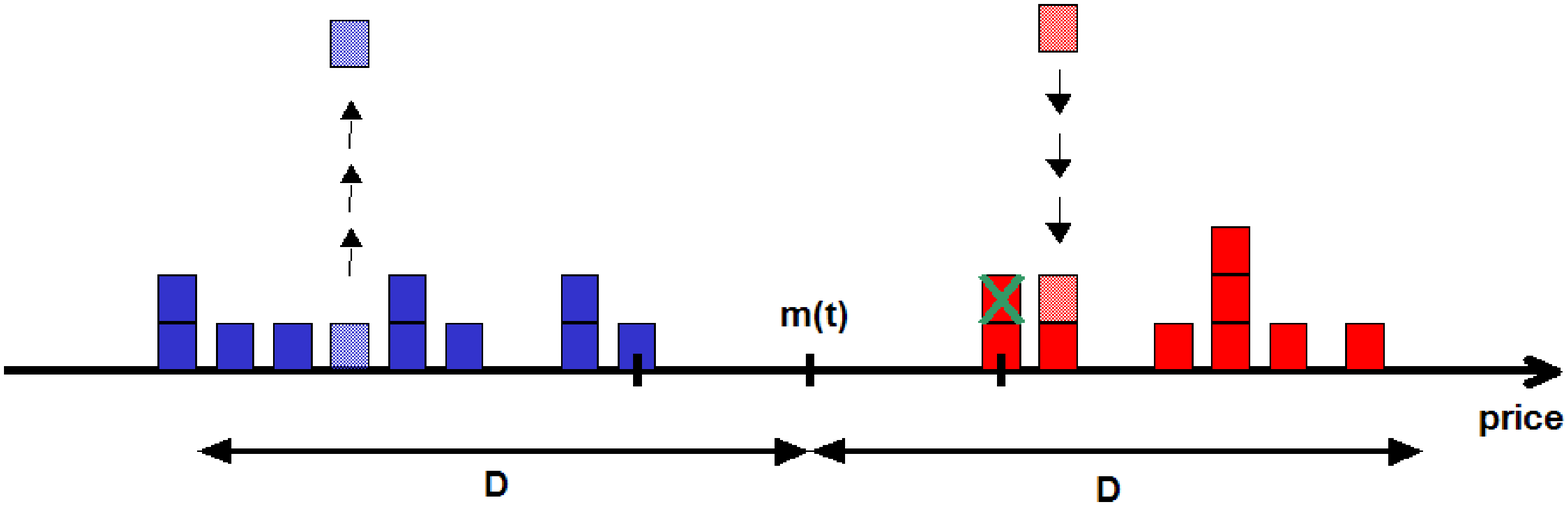}
}
}
\caption{(Color online) Scheme of the dynamics of the
zero intelligence model.}
\label{fig:model_scheme}
\end{figure}

The three rates add up to one: $P_{LO}+P_{MO}+P_{C}=1$
(in other words we study the model in event time).
In the beginning of the simulation there is a long warming up period,
when orders are randomly placed in the book, in order not to have
spurious results due to fluctuations and an empty book.
Similarly to the empirical analysis, we denote the mid-price
at time $t$ by $m_t$ and the bid-ask spread at time $t$ by $S_t$.
Figure \ref{fig:model_scheme} shows the scheme of the
order flow mechanism. Since in our model there are no crossing limit
orders, in order to have a stationary number of orders we set $P_{LO}=0.5$.

We generate large price changes manually, by clearing out all limit
orders on one side of the book\footnote{This method is clearly not realistic.
However, this is the only way we could think to generate large price changes in a zero
intelligence model, i.e., when price jumps are not governed by sudden changes
in the agents perception, strategy and herding behaviour.
Since we are
interested in the relaxation after the jump, we believe that this way of
generating the large price changes is satisfactory.} in the interval
$[b_{t}-J,b_{t}]$ for price drops and
$[a_{t},a_{t}+J]$ for price jumps, where $J$ is a
parameter of the model (and $b_t$ and $a_t$ are the best bid and best
ask respectively), this way the jump in the mid-price is $J/2$.

As we have mentioned, with such a simple model
we have to confine ourselves to the study of the bid-ask spread and
the volatility, i.e., variables whose dynamics may be studied
in a model without adaptation rules and agents' intelligence
(unlike other variables, like the activity, that are more closely
related to changes in agents' perception and thus cannot be analysed
in a simple zero intelligence model).

\subsection{Numerical results}\label{numerics}
For usual market dynamics the rate of different orders are roughly
$P_{LO}=0.5$, $P_{MO}=0.16$ and $P_{C}=0.34$ \cite{farmer2004}\footnote{Note that
these rates are defined for \textit{effective} limit orders and
\textit{effective} market orders, i.e. all orders that lead to an
immediate execution are regarded as market
orders. Since we do not have crossing limit orders in our simulation, it
is right to use these values. }. In this section we present the results of
our numerical simulations using the above empirical probabilities. The
parameters of the simulation are the following:

\begin{itemize}
\item $D=1000$
\item $J=1000$
\item the frequency of large price jumps was $f=(5*10^{4})^{-1}$
\item the length of the simulation was $5*10^6$.
\end{itemize}

\begin{figure*}[htb!]%
\centering
\subfigure[][]{%
\label{fig:model_relax-volat}%
\includegraphics[width=0.35\textwidth,angle=-90]{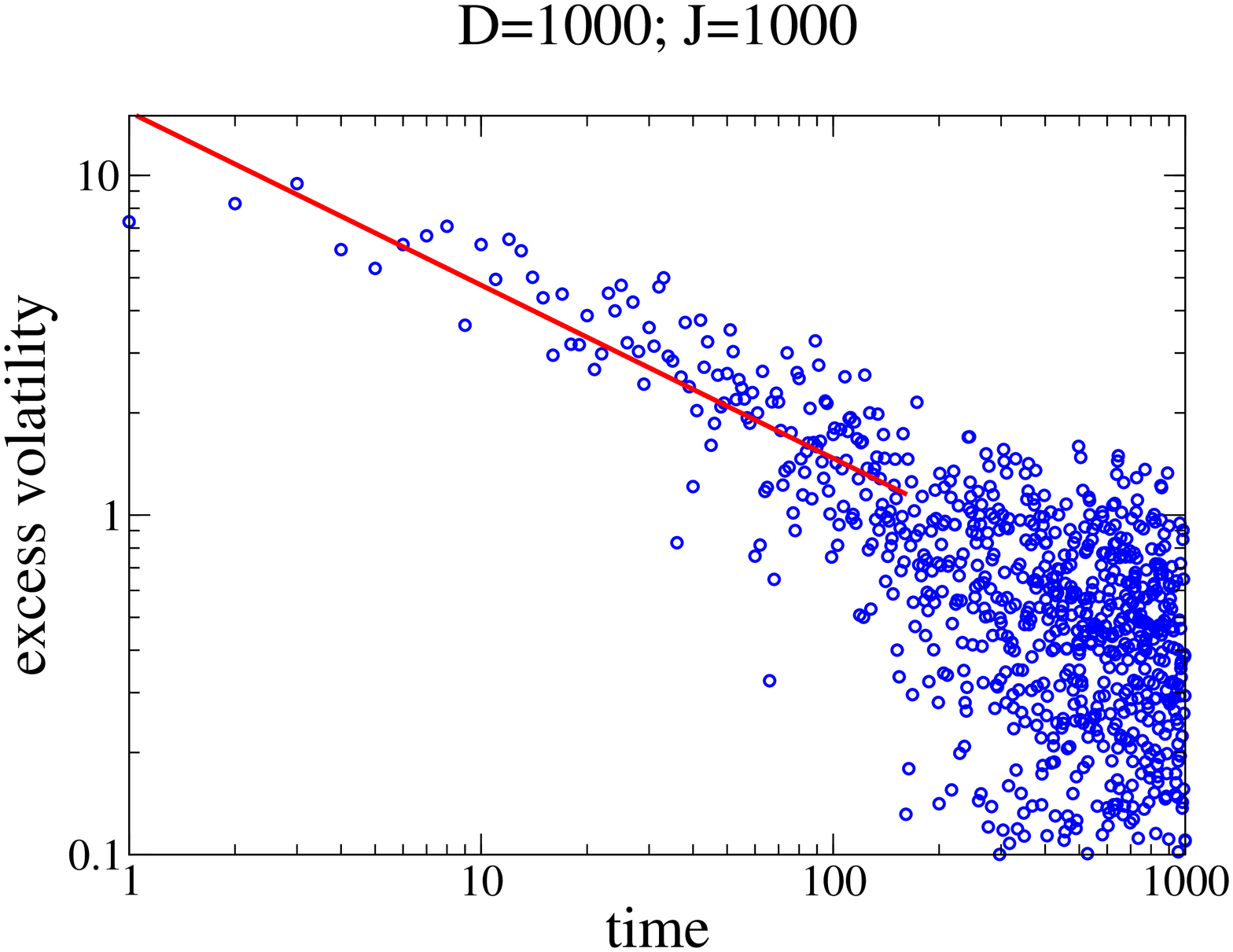}}%
\hspace{8pt}%
\subfigure[][]{%
\label{fig:model_relax-spr}%
\includegraphics[width=0.35\textwidth,angle=-90]{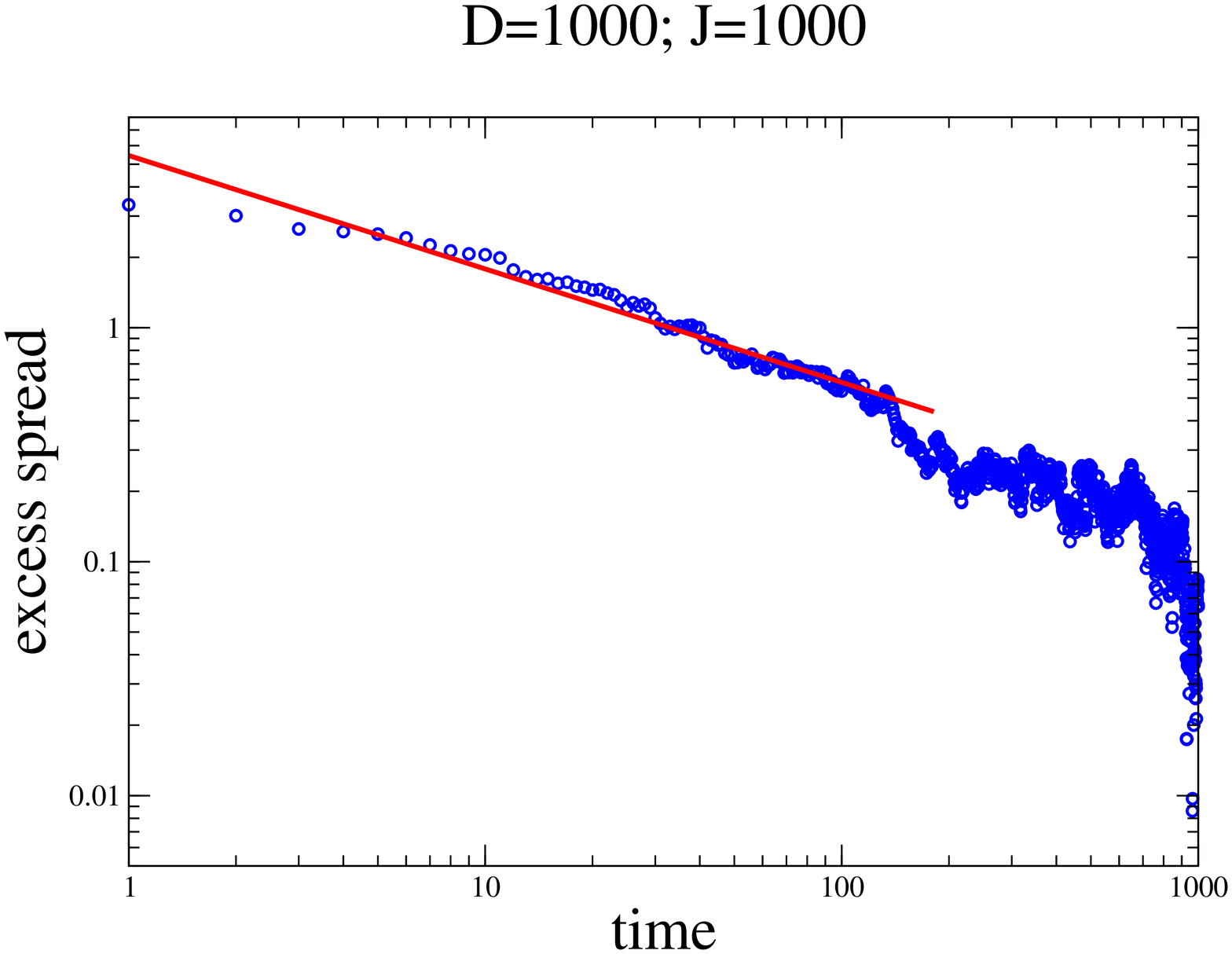}}\\
\caption[A set of four subfigures.]{(Color online) Relaxations after price jumps in
the numerical model, in case of $P_{LO}=0.5$, $P_{MO}=0.16$ and $P_{C}=0.34$.
We also show the power law fits of the relaxations.
\subref{fig:model_relax-volat} relaxation of the volatility, $0.50\pm 0.04$;
\subref{fig:model_relax-spr} relaxation of the bid-ask spread, $0.48\pm 0.01$;
Both exponents are close to 0.5. Similarly to
empirical results, the variation in the volatility is much stronger than
in the bid-ask spread.}
\label{fig:model_relax}%
\end{figure*}

Figure \ref{fig:model_relax} shows the average decay of the
volatility and the bid-ask spread after large price changes in the model.
The plots are
created in the manner as the figures in Section \ref{empirical}: Time
zero is the moment of the event and the y-axis shows the relative
dynamics compared to stationary market periods, averaged over 100
events. The decays are qualitatively similar to what we have seen for the
empirical data. The short time relaxations (up to roughly 100 simulation steps)
can be described by power laws, with
exponents of roughly 0.5 ($0.50\pm 0.04$ for the volatility and $0.48\pm 0.01$
for the bid-ask spread).
However, there is a difference between the empirical and numerical exponents.
This suggests that our simple model is not able to entirely reproduce
the relaxations. It seems that part of the slow relaxation can be generated
in the model, however the difference between the exponents $0.4$ and $0.5$
is important and we believe that the discrepancy is due to the
behaviour of the agents, that can not be captured in the zero intelligence model.
It is interesting that the exponent of the decay in the
volatility and in the bid-ask spread seem to be very close to each other,
similarly as in the case of the empirical data.
Also similarly to the empirical data, the peak in the volatility is
smaller than that in the bid-ask spread.

\subsection{Analytical treatment}\label{analytics}
As we have seen, the bid-ask spread shows a slow relaxation
both for empirical data and simulations. To
understand if it is really a critical relaxation and what determines
the exponent, we tried to treat the model analytically.

\subsubsection{A limit case}\label{limit}
The model can be treated analytically in the case of
$P_{MO}=0$, i.e. when only limit order placing and cancelation
determine the flow ($P_{LO}=P_{C}=0.5$), using some simple assumptions.

Mathematically the bid-ask spread in this limit case can be understood as the
time evolution of the minimum of numbers distributed uniformly
on a finite interval (except for very short times). This tells
us that the bid-ask spread will decay according to a power law with
unit exponent, clearly showing that the limit case is unrealistic.
Nevertheless we discuss it briefly as we are going to use these
results in the following.

When describing the time evolution of the bid-ask spread we use a simplifying assumption: Since the
probability of the best order to be canceled on either side is very
low, we can assume that cancelations do not alter the value of the
bid-ask spread.  With the above assumption, the expectation of the change in
bid-ask spread from one time step to another can be written in the following
way:

\begin{eqnarray}\label{eq:model1}
\mathbb{E}(\Delta S_t)=P_{C}\cdot 0+P_{LO}\Big( \frac{D-\frac{S_t}{2}}{D}\cdot 0-\frac{1}{D}\sum_{k=1}^{\frac{S_t}{2}}k\Big),
\end{eqnarray}

\noindent where the first term on the right hand side assumes that
cancelations do not change the bid-ask spread, and the second term describes
the effect of a new limit order arriving: If it falls outside the bid-ask spread
it does not change the bid-ask spread, if it falls inside the bid-ask spread with
a distance $k$ from the same best price, it decreases the bid-ask spread exactly by $k$.
Summing up the right side of Eq. \ref{eq:model1} we get

\begin{eqnarray}\label{eq:model1b}
\mathbb{E}(\Delta S_t)=-P_{LO}\Big( \frac{S_{t}^{2}}{8D}+\frac{S_t}{4D}\Big).
\end{eqnarray}

Using Equation \ref{eq:model1b}, in case of $P_{LO}=0.5$ the
expected value of the bid-ask spread can be written in the following recursive
formula:

\begin{eqnarray}\label{eq:model2}
\mathbb{E}(S_{t+1})=\mathbb{E}(S_{t})+\mathbb{E}(\Delta S_{t})=\nonumber \\
=\mathbb{E}(S_{t})\Big(1-\frac{1}{8D}\Big)-\frac{\mathbb{E}(S_{t})^{2}}{16D}.
\end{eqnarray}

The above recursive formula is a mean field theory that can be used
to describe the relaxation
when knowing the initial bid-ask spread after the event. The formula fits
the numerical results very well reproducing the asymptotic power
law with exponent very close to 1.

\subsubsection{General case}\label{general}
We are most interested in the general case when the probability of market
orders is finite. For $P_{MO}\neq 0$, the
change of the bid-ask spread is the following:

\begin{eqnarray}\label{eq:model3}
\mathbb{E}(\Delta S_t)=P_{C}\cdot 0+P_{LO}\Big( \frac{D-\frac{S_t}{2}}{D}\cdot 0-\frac{1}{D}\sum_{k=1}^{\frac{S_t}{2}}k\Big)+\nonumber \\
+P_{MO}\cdot g^{(1)}_{t}=-P_{LO}\Big[\frac{S_{t}^{2}}{8D}+\frac{S_t}{4D}\Big]+P_{MO}\cdot g^{(1)}_{t},
\end{eqnarray}

\noindent where the first term assumes again that cancelations do not
change the bid-ask spread, the second term gives the expected change in case of
a limit order arriving (similarly to the limit case, Eq. \ref{eq:model1}) and the
last term stands for a market order arriving, increasing the
bid-ask spread exactly by the size of the first gap, $g^{(1)}_{t}$ 
(the gap is defined in the same way as in the empirical case). As we can
see, when having market orders, not surprisingly we have to account
for the gaps as well. The expectation for the gap can be given by the
following equation:

\begin{eqnarray}\label{eq:model4}
\mathbb{E}(g^{(1)}_{t})=P_{C}\cdot g^{(1)}_{t-1}+\nonumber \\
+P_{LO}\Bigg[ \frac{D-\frac{S_{t}}{2}-g^{(1)}_{t-1}}{D}g^{(1)}_{t-1}+\frac{1}{D}\sum_{k=1}^{g^{(1)}_{t-1}}k+\nonumber \\
+\frac{1}{D}\sum_{k=1}^{\frac{S_{t-1}}{2}}k\Bigg]+P_{MO}\cdot g^{(2)}_{t-1},
\end{eqnarray}

\noindent where $g^{(2)}$ stands for the second gap in the book, i.e. the price
difference between the second best and third best order on the same
side of the book. In Equations \ref{eq:model3} and \ref{eq:model4}
we neglected the probability of cancelations changing the bid-ask spread or
the first gap. As we can see, in the expected value of the first gap,
a term containing the second gap occurs. This is the general case for
all gaps that is, when writing up $g^{(n)}$, it will contain a term
depending on $g^{(n+1)}$. We have to find a closure for this infinite
hierarchy of equations, so we do the following.
First, we estimate the relation between the first and the second gaps
for the equilibrium state of the system (the value of the second gap,
as a function of the first gap).
Second, we assume this relation to be constant also for the
relaxation period.
By equilibrium state, here we mean the long time behaviour of the system
without large price jumps.
We denote the equilibrium values of the bid-ask spread, the
first gap and the second gap by $\sigma$, $\gamma^{(1)}$ and $\gamma^{(2)}$
respectively. In case of stationarity, $\mathbb{E}(\Delta S_t)=0$,
thus Equation \ref{eq:model3} becomes:

\begin{eqnarray}\label{eq:model5}
P_{LO}\Big[\frac{\sigma^{2}}{8D}+\frac{\sigma}{4D}\Big]=P_{MO}\cdot\gamma^{(1)}.
\end{eqnarray}

\noindent Equation \ref{eq:model4} becomes

\begin{eqnarray}\label{eq:model6}
\gamma^{(1)}=P_{C}\cdot\gamma^{(1)}+\nonumber \\
+P_{LO}\Bigg[ \frac{D-\frac{\sigma}{2}-\gamma^{(1)}}{D}\gamma^{(1)}+\frac{1}{D}\sum_{k=1}^{\gamma^{(1)}}k+\nonumber \\
+\frac{1}{D}\sum_{k=1}^{\frac{\sigma}{2}}k\Bigg]+P_{MO}\cdot\gamma^{(2)}.
\end{eqnarray}

\noindent Combining Equations \ref{eq:model5} and \ref{eq:model6}, we get
the following formula:

\begin{eqnarray}\label{eq:model7}
\frac{\gamma^{(2)}}{\gamma^{(1)}}=1+\frac{1}{2D}\frac{P_{LO}}{P_{MO}}\Big[ \sigma+\gamma^{(1)}-1\Big]-2P_{LO}.
\end{eqnarray}

\noindent Knowing that $P_{LO}=0.5$ in the model, we get

\begin{eqnarray}\label{eq:model8}
\frac{\gamma^{(2)}}{\gamma^{(1)}}=\frac{1}{2D}\frac{P_{LO}}{P_{MO}}\Big[ \sigma+\gamma^{(1)}-1\Big].
\end{eqnarray}

The above relation between the first and second gap can be assumed to
be true for the relaxation process, since the second gap has only a
minor role in case of decreasing bid-ask spread. Introducing Equation
\ref{eq:model8} into Equation \ref{eq:model4}, we get a recursive formula
for the size of the first gap, and through that for the size of the
bid-ask spread. To be able to use the relations, we need to know the expected
value of the stationary bid-ask spread, $\sigma$. This we know from the decay
of the bid-ask spread: The value, of the first point (the relative opening of
the bid-ask spread) is approximately $(\sigma+J)/\sigma$.

\begin{figure}[htb!]
\begin{center}
\resizebox{0.75\columnwidth}{!}{%
  \includegraphics[width=0.25\textwidth,angle=-90]{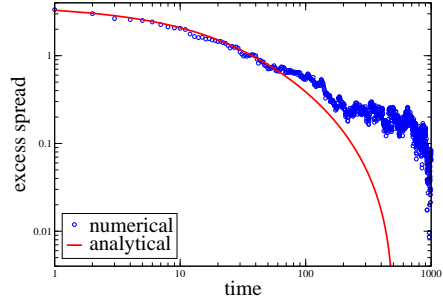}
}
\caption{(Color online) The relaxation of the bid-ask spread after
large price changes in case of $P_{MO}=0.16$. The black circles show
the numerical values, the red line shows the analytical values.}
\label{fig:model_spread_P016_fit}
\end{center}
\end{figure}

Figure \ref{fig:model_spread_P016_fit}
shows the comparison of the numerical and the analytical results for
the relaxation of the bid-ask spread for $P_{MO}=0.16$.

The analytical formula seems to describe the relaxation of the bid-ask spread for
short times but not long times. We can see that the agreement between the
analytical and numerical curves only hold for roughly the first 100 simulation
steps in the relaxation. This is due to the fact that on long times the ``mean
field'' assumption ignoring the actual dynamics of the second gap gives rise
to larger errors.

\section{Conclusions}\label{conclusions}

We have studied the dynamics of several measures of the order book
after large price changes.
For the change in the volatility, the bid-ask spread, the limit order placing
and cancelation activity, the bid-ask imbalance, the number of queuing
orders in the book we found strong variation at the moment of the event and
slow relaxation in the post-event period. Specially, in case of the volatility,
the bid-ask spread and the activities we found a relaxation very similar to a power
law, with exponents close to 0.4.

Analysing the rates of limit orders, market orders and cancelations around
large events, we did not find strong variation, showing that strategy of
traders in choosing their type of order does not vary much.

To deeper understand the similar slow relaxations found empirically,
we constructed a zero intelligence multi agent model for the order flow.
The model is essentially a deposition-evaporation model with market orders
added. The large price changes were generated manually. We found that the
simple model was able to reproduce the relaxation of volatility and bid-ask
spread qualitatively. The relaxations in the model were slow, similar to a
power law with exponents very close to each other, similarly to the case of
empirical data. The value of the exponents found was roughly 0.5 both for
the volatility and the bid-ask spread, slightly higher than empirically. The
ratio of the peak in the volatility and in the bid-ask spread was also similar to the
one found empirically. Consequently, though we find that the relaxations
are slower in real markets than in the simulations, the values suggest that
the overall character of the slow relaxations can be explained in the framework
of the zero intelligence model without assuming strategic behaviour of
agents.
However the difference between the exponents $0.4$ and $0.5$
is important and we believe that the discrepancy is due to the
behaviour of the agents, that can not be captured in the zero intelligence model.

We gave an analytic solution for the relaxation of the bid-ask spread in
the model for the limit case of $P_{MO}=0$ and for the short term relaxation in
case of arbitrary $P_{MO}$.

\section*{Acknowledgments}
Support by OTKA Grants T049238
and K60456 is acknowledged. J.D.F. acknowledges support from Barclays Bank, Bill Miller
and NSF grant HSD-0624351. Any opinions, findings and conclusions
or recommendations expressed in this material are
those of the authors and do not necessarily reflect the views of
the National Science Foundation.


\begin{thebibliography}{99}

\bibitem{omori1894} F. Omori, \emph{On the after-shocks of earthquakes},
J. Coll. Sci. Imp. Univ. Tokyo, \textbf{7}, 111--200 (1894)

\bibitem{chamberlin1996} R. V. Chamberlin,
\emph{Universalities in the primary response of condensed matter}
Europhysics Letters, \textbf{33}, 545 (1996)

\bibitem{bouchaud1992} J.-P. Bouchaud, \emph{Weak ergodicity breaking and aging in disordered systems},
J. Phys. I, \textbf{2}, 1705--1713 (1992)

\bibitem{zapperi1997} S. Zapperi, A. Vespignani, H. E. Stanley,
\emph{Plasticity and avalanche behaviour in microfracturing phenomena},
Nature, \textbf{388}, 658 (1997)

\bibitem{johansen2000} A. Johansen, D. Sornette,
\emph{Download relaxation dynamics on the WWW following newspaper publication of URL},
Physica A, \textbf{276}, 338-345 (2000)

\bibitem{abe2003} S. Abe, N. Suzuki,
\emph{Omori's law in the Internet traffic},
Europhysics Letters, \textbf{61}, No. 6, 852-855 (2003)

\bibitem{lillo2003} F. Lillo, R.N. Mantegna, \textit{Power law relaxation in a
complex system: Omori law after a financial market crash}, Physical
Review E, \textbf{68}, 016119 (2003)

\bibitem{zawadowski2004} A.G. Zawadowski, J. Kert\'esz, G. Andor,
\emph{Large price changes on small scales},
Physica A \textbf{344}, 221-226 (2004)

\bibitem{zawadowski2006} A.G. Zawadowski, G. Andor, J. Kert\'esz,
\textit{Short-term market reaction after extreme price changes of
liquid stocks}, Quantitative Finance, \textbf{6}, 283--295 (2006)

\bibitem{farmer2004} J. D. Farmer, L. Gillemot, F. Lillo, S. Mike, A. Sen,
\textit{What really causes large price changes?}, Quantitative
Finance, \textbf{4}, 383--397 (2004)

\bibitem{weber2006} P. Weber, B. Rosenow, \textit{Large stock price changes:
volume or liquidity}, Quantitative Finance, \textbf{6}, 7--14 (2006)

\bibitem{ponzi2006} A. Ponzi, F. Lillo, R.N. Mantegna, \textit{Market reaction
to temporary liquidity crises and the permanent market impact} preprint:
http://arxiv.org/abs/physics/0608032 (2006).

\bibitem{joulin2008} A. Joulin,
A. Lefevre, D. Grunberg, J.-P. Bouchaud, \textit{Stock price jumps:
news and volume play a minor role}, preprint:
http://arxiv.org/abs/0803.1769 (2008)

\bibitem{eisler2007} Z. Eisler, J. Kert\'esz, F. Lillo, R. N. Mantegna,
\emph{Diffusive behavior and the modeling of characteristic times in limit order executions},
preprint: http://lanl.arxiv.org/abs/physics/0701335 (2007)

\bibitem{lse} http://www.londonstockexchange.com

\bibitem{zawadowski_private}
A.G. Zawadowski, private communication

\bibitem{weber2007} P. Weber, F. Wang, I. Vodenska-Chitkushev, S. Havlin, H. E. Stanley,
\emph{Relation between volatility correlations in financial markets and Omori
processes occurring on all scales}, Physical Review E, \textbf{76}, 016109 (2007)

\bibitem{stigler1963} G. J. Stigler, \emph{Public Regulation of the Securities Markets},
Journal of Business, \textbf{37}, 117, (1963)

\bibitem{bak1997} P. Bak, M. Paczuski, M. Shubik, \emph{Price Variations in a
Stock Market With Many Agents},
Physics A, \textbf{246}, 430 (1997)

\bibitem{maslov2000} S. Maslov, \textit{Simple model of a limit order-driven market},
Physica A, \textbf{278}, 571-578 (2000)

\bibitem{challet2001} D. Challet, R. Stinchcombe, \emph{Analyzing and modelling 1+1d markets},
Physica A, \textbf{300}, 285 (2001)

\bibitem{willmann2002} R. D. Willmann, G. M. Sch\"utz, D. Challet,
\emph{Exact Hurst exponent and crossover behavior in a limit order market model},
Physica A, \textbf{316}, 430 (2002)

\bibitem{muchnik2003} L. Muchnik, F. Slanina, S. Solomon, \emph{The interacting gaps model:
reconciling theoretical and numerical approaches to limit-order models}, Physica A,
\textbf{330}, 232 (2003)

\bibitem{daniels2003} M.G. Daniels, J.D. Farmer, L. Gillemot, G. Iori, E. Smith,
\textit{Quantitative Model of Price Diffusion and Market Friction Based on Trading
as a Mechanistic Random Process}, Physical Review Letters, \textbf{90}(10), 108102 (2003)

\bibitem{smith2003} E. Smith, J.D. Farmer,
L. Gillemot, S. Krishnamurthy, \textit{Statistical theory of the
continuous double auction}, Quantitative Finance, Vol. 3. 481-514
(2003)

\end{thebibliography}
\end{document}